\documentclass{vgtc}                          

\ifpdf
  \pdfoutput=1\relax                   
  \usepackage{graphicx}                
  \DeclareGraphicsExtensions{.pdf,.png,.jpg,.jpeg} 
\else
  \usepackage{graphicx}                
  \DeclareGraphicsExtensions{.eps}     
\fi%

\graphicspath{{figures/}{pictures/}{images/}{./}} 

\usepackage{microtype}                 
\PassOptionsToPackage{warn}{textcomp}  
\usepackage{textcomp}                  
\usepackage{mathptmx}                  
\usepackage{times}                     
\usepackage{cite}                      
\usepackage{tabu}                      
\usepackage{booktabs}                  
\usepackage{amsmath}
\usepackage{float}
\usepackage[caption=false]{subfig}
\usepackage{lscape}
\usepackage{comment}
\usepackage{amsbsy}
\usepackage{stackengine}
\usepackage{dblfloatfix}
\usepackage{afterpage}
\usepackage[ruled,vlined]{algorithm2e}

\onlineid{3593}

\vgtccategory{Research}

\vgtcinsertpkg

\title{A Quality Metric for Symmetric Graph Drawings}

\author{Amyra Meidiana\thanks{e-mail: amei2916@uni.sydney.edu.au}\\
\scriptsize \centering University of Sydney %
\and Seok-Hee Hong\thanks{e-mail: seokhee.hong@sydney.edu.au}\\
\scriptsize \centering University of Sydney %
\and Peter Eades\thanks{e-mail: peter.eades@uni.sydney.edu.au}\\
\scriptsize \centering University of Sydney
\and Daniel Keim\thanks{e-mail: keim@uni-konstanz.de}\\ %
     \scriptsize \centering University of Konstanz}

\abstract{Symmetry is an important aesthetic criteria in graph drawing and network visualisation. Symmetric graph drawings aim to faithfully represent \textit{automorphisms} of graphs as geometric symmetries in a drawing. 

In this paper, we design and implement a framework for quality metrics that measure symmetry, that is, how faithfully a drawing of a graph displays automorphisms as geometric symmetries. The quality metrics are based on \textit{geometry} (i.e. Euclidean distance) as well as mathematical \textit{group theory} (i.e. orbits of automorphisms).

More specifically, we define two varieties of symmetry quality metrics: (1) for displaying a single automorphism as a symmetry (\textit{axial} or \textit{rotational}) and (2) for displaying a  group of automorphisms (\textit{cyclic} or \textit{dihedral}).  We also present algorithms to compute the symmetric quality metrics in \(O(n \log n)\) time for rotational symmetry and axial symmetry.

We validate our symmetry quality metrics using deformation experiments. We then use the metrics to evaluate a number of established graph drawing layouts to compare how faithfully they display automorphisms of a graph as geometric symmetries.  } 

\CCScatlist{ 
  \CCScat{Human-centered computing}{Visu\-al\-iza\-tion}{Visualization design and evaluation methods}{}
}

\begin{document}


\firstsection{Introduction}

\maketitle

Graph drawing aims to construct a visually-informative drawing of an
abstract graph in the plane. Symmetry is one of the most important
aesthetic criteria that represent the structure and properties of
a graph visually~\cite{battista1998graph}. Symmetric graph drawings aim to faithfully represent automorphisms of graphs as geometric symmetries in the drawing. Also, a symmetric drawing of a graph enables an understanding of the entire graph to be built up from that of a smaller subgraph.

Symmetric drawings of a graph $G$ are clearly related to the automorphisms of $G$, and algorithms for constructing symmetric drawings have two steps:\\
\\
{\em Step 1. Find the ``appropriate'' automorphisms}, and \\
{\em Step 2. Draw the graph displaying these automorphisms as symmetries.}\\

The problem of determining whether a graph has a nontrivial automorphism is {\em automorphism
complete}~\cite{lubiw1981}. However, the problem of determining whether a graph has a nontrivial {\em geometric} automorphism is {\em NP-complete}~\cite{manning1992geometric}. Linear time algorithms to construct symmetric drawings for restricted classes of graphs exists (e.g. trees~\cite{manning1988fast}, outerplanar graphs~\cite{manning1992fast}, series-parallel digraphs~\cite{hong2000drawing}), and planar graphs 
~\cite{hong2006linear,hong2003symmetric,hong2005drawing,hong2006drawing}. For general graphs, heuristics~\cite{lipton1985method, defraysseix1999heuristic} and exact algorithms are available
~\cite{abelson2007geometric,buchheim2003}.

It is important to ensure that a symmetric graph drawing accurately represent the automorphisms of the underlying graph to the greatest extent possible. However, existing symmetry detection and quality metrics for graph drawings do not focus on this comparison between detected geometric symmetry in a graph drawing and the automorphisms of the graph.

In this paper, we design and implement a framework for quality metrics for graph drawings that measure symmetry, that is, how faithfully a drawing of a graph displays automorphisms as geometric symmetries. The quality metrics are based on geometry (i.e., Euclidean distance) as well as mathematical group theory (i.e., orbits of automorphisms).

More specifically, we present the following contributions:

\begin{enumerate}
    \item We design and implement a framework for a quality metric for symmetric graph drawing, which measures the \textit{symmetry quality} of a drawing of a graph based on the comparison between the symmetry of the drawing and the automorphism of the graph. We define two varieties of the symmetry quality metrics, one to measure how well a graph drawing displays a single automorphism as a symmetry (rotational or axial) and one to measure how well a graph drawing displays a group of automorphism simultaneously (cyclic or dihedral groups).
    \item We present algorithms to compute the symmetry quality metrics in \(O(n \log n)\) time for rotational and axial symmetry and \(O(kn \log n)\) for automorphism groups, where \(n\) is the number of vertices in the graph and \(k\) is the size of the automorphism group.
    \item We validate the single automorphism detection version of the symmetry quality metrics through deformation experiments of graph drawings, showing that the scores computed by our metric decrease when the drawings are distorted further from exact symmetry.
    \item We validate the automorphism group detection version of the symmetry quality metrics through comparing drawings displaying different groups of automorphisms,  showing that our metric effectively captures the difference in symmetry quality between drawings that display different numbers of automorphisms as symmetries.
    \item We use our metric to compare a number of established graph drawing layouts to compare how faithfully they display automorphisms of a graph as geometric symmetries. We confirm the effectiveness of the \textit{concentric circles} layout in displaying a graph's automorphisms as symmetries.
\end{enumerate}

\section{Related Works}

\subsection{Symmetries and geometric automorphisms}

An {\em automorphism} of a graph is a permutation of the vertices that preserves the adjacency of each vertex. Suppose that a permutation group $A$ acts
on a set $V$. We say that $\phi \in A$ {\em fixes} $v \in V$ if
$\phi(v) = v$; if $\phi$ fixes $v$ for every $\phi \in A$ then $A$
{\em fixes} $v$. If $V' \subseteq V$ and $\phi(v') \in V'$ for all
$v' \in V'$ then $\phi$ {\em fixes} $V'$ ({\em setwise}). A subset
$V'$ of $V$ is an {\em orbit} of $A$ if, for each $u, v \in V'$,
there is $\phi \in A$ such that $\phi (u) = v$, and $V'$ does not
contain a nonempty subset with this property. 

It is important to use a rigorous model, introduced by Eades and Lin~\cite{eades2000spring}, for the intuitive concept
of symmetry display. 

The symmetries of a set of points in the plane (such as a two
dimensional graph drawing) form a group called the {\em symmetry
group} of the set. A symmetry $\sigma$ of a drawing $D$ of a graph
$G$ {\em induces} an automorphism of $G$, in that the restriction
of $\sigma$ to the points representing vertices of $G$ is an
automorphism of $G$. A drawing $D$ of a graph $G$ {\em displays}
an automorphism $\phi$ of $G$ if there is symmetry $\sigma$ of $D$
which induces $\phi$. The symmetry group of a graph drawing
induces an automorphism group of the graph. An automorphism group
$A$ of a graph $G$ is a {\em geometric automorphism group} if
there is a drawing of $G$ which displays every element of $A$.

A non-trivial symmetry of a finite set of points in the plane is
either a rotation about a point or a reflection about a line. A
geometric automorphism is a {\em rotational automorphism}
(respectively {\em relectional/axial automorphism}) if it is induced by a rotation (respectively reflection).
Eades and Lin~\cite{eades2000spring} showed that that a nontrivial geometric automorphism group is one
of three kinds:

\begin{enumerate}
\item a group of size 2
        generated by an axial automorphism;\\
\item a {\em cyclic group} of size $k$
        generated by a rotational automorphism;\\
\item a {\em dihedral group} of size
        $2k$ generated by a rotational
        automorphism of order $k$ and an axial automorphism.\\
\end{enumerate}
        
\subsection{Symmetric Graph Drawing}

In general, determining whether a graph can be drawn symmetrically in two dimensions is NP-complete~\cite{lubiw1981}.
Exact algorithms based on Branch and Cut~\cite{buchheim2003} and group theory~\cite{abelson2007geometric} are available. 

Linear-time algorithms are available for symmetric drawings of limited classes of graphs, such as trees~\cite{manning1988fast}, outerplanar graphs~\cite{manning1992fast}, and series-parallel digraphs~\cite{hong2000drawing}. Linear-time algorithms have also been presented for maximally symmetric drawings of triconnected~\cite{hong2006linear}, biconnected~\cite{hong2003symmetric}, oneconnected~\cite{hong2005drawing}, and disconnected planar graphs~\cite{hong2006drawing}.
For a survey on symmetric drawings of graphs in two dimensions, see~\cite{eades2013detection}.

\subsection{Graph Drawing Symmetry Quality Metrics}

Purchase~\cite{purchase2002metrics} defined a metric measuring the symmetry of a graph drawing by checking, for each pair of vertices, whether there is a symmetric subgraph around the pair, calculating a weighted symmetry value of the symmetric subgraph if it exists, and adding the weighted symmetry value of all symmetric subgraphs. However, this metric has a best runtime of \(O(n^5)\) and only considers axial symmetry.

Klapaukh et al~\cite{klapaukh2018} defined a metric which detects rotational, axial, and translational symmetry in a node-link graph drawing, using methods adapted from computer vision. Taking as input an image of a drawing of a graph, this method detects symmetries by converting edges detected in the image into line vectors and then computing the symmetry quality for each line and each pair of lines, giving the runtime as \(O(m^2)\) time in the number of edges, with the worst case of \(O(n^4)\) time in the number of vertices for dense graphs.

\subsection{Geometric Symmetry Detection}

Optimal algorithms for exact symmetry detection for two-dimensional point sets and polygons run in \(O(n \log n)\) time for detecting rotational or axial symmetry in two-dimensional point sets~\cite{wolter1985optimal, alt1988congruence}. These algorithms work by sorting the points by their angle around and distance from a centroid, then finding a palindromic sequence of angles and distances (for axial symmetry) or a subsequence that is repeated within the full sequence (for rotational symmetry). However, these algorithms only give a binary answer to whether a point set displays a symmetry and are not suitable to quantify the approximate symmetry of a point set.

Zabrodsky et al~\cite{zabrodsky1995symmetry} introduces the \textit{symmetry distance} method, providing another method to quantify the approximate symmetry of a geometric object. This method takes as input a point set \(P\) and a symmetry \(\sigma\) to be checked and computes a symmetric point set \(P_\sigma\) realizing the input symmetry. This is done through a ``folding'' transform method that ``folds'' points in the same orbit, averages their positions, and ``unfolds'' them into a symmetric configuration. This method minimizes the Euclidean distance between each point in \(P\) and its image in \(P_\sigma\), and the symmetry distance is given as the average of these Euclidean distances for every point in \(P\).

\section{Symmetry Quality Metric Framework}

\begin{figure}[ht]
\centering
\includegraphics[width=1\columnwidth]{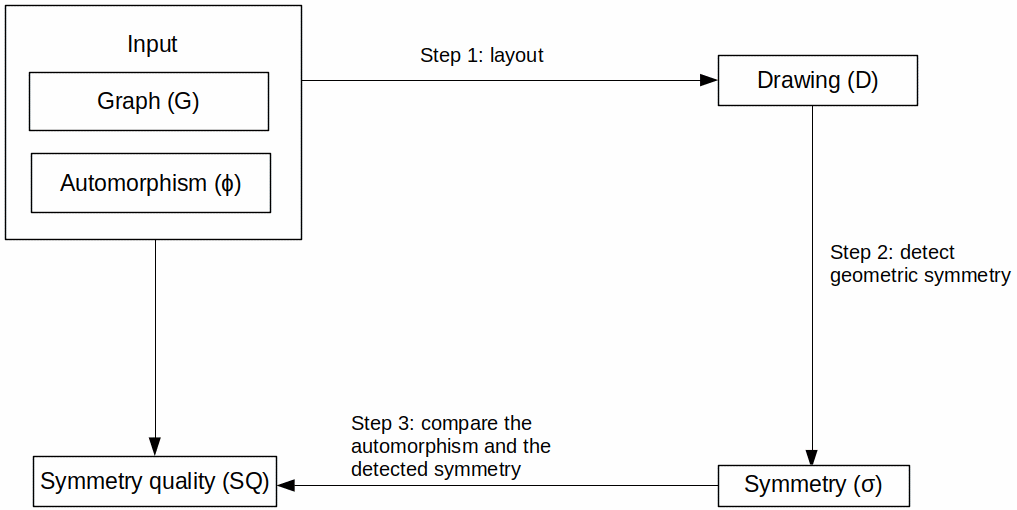}
\caption{Framework for the symmetry quality metric.}
\label{fig:symframework}
\end{figure}

We propose a new quality metric for graph visualization, the \textit{symmetry quality metric}, for measuring how well the drawing of a graph displays selected geometric automorphisms of the underlying graph. Our metric is a faithfulness metric comparing the geometric symmetry detected in a drawing of a graph with the graph's automorphisms, the ground truth information, unlike existing metrics which only attempts to detect symmetry from the drawing. 
Figure \ref{fig:symframework} summarizes the framework used for our proposed metric.

Let \(G\) be a graph and \(\phi\) be a automorphism of \(G\). It is important that \(\phi\) be a \textit{geometric} automorphism of \(G\), as otherwise it is impossible to display it as a symmetry of a drawing of \(G\)~\cite{eades2000spring}. The framework computes the symmetry quality score using the following steps:
\\
\\
\textbf{Framework: Symmetry Quality Metric}
\begin{description}
    \item [Step 1:] Apply a layout algorithm to \(G\) to obtain a graph drawing \(D\), which provides geometric positions for each vertex in \(G\).
    \item [Step 2:] Detect a geometric symmetry of \(D\), obtaining symmetry \(\sigma\).
    \item [Step 3:] Compare \(\sigma\) to \(\phi\) to compute the symmetry quality metric.
\end{description}

\subsection{Symmetry Quality Metric for a Single Automorphism}

Given a drawing of a graph with an exact symmetry, a geometric symmetry detection algorithm can be used to detect a symmetry of the drawing, which induces an automorphism of the graph, and this result can be compared to the input automorphism. However, in practice, automatic graph layout algorithms may produce drawings that are not perfectly symmetric. We therefore define a refinement that uses \textit{approximate symmetry detection} to quantify how far the drawing is from an exact symmetry that displays an input automorphism.

The approximate symmetry of a point set \(P\) can be defined as a Euclidean distance between it and a symmetric point set \(P_\sigma\). In this refinement of our metric, for each orbit of \(\phi\), we compute the Euclidean distance needed to transform the position of its points in \(D\) to perfect symmetry, and use this distance to compute the symmetry quality.

We define two different formulas to compute the symmetry quality metric, \(SQ\), given the number of symmetric orbits and the Euclidean distance to exact symmetry for the asymmetric orbits. The first formula gives equal weight to the proportion of symmetric orbits and the average Euclidean distance to perfect symmetry of asymmetric orbits:
\begin{equation}
SQ_1 = \frac{1}{2} \left(\frac{|O_{sym}|}{|O|} + \frac{1}{|O_{asym}|} \sum_{k=1}^{|O|} sd(o_k)\right)
\label{eq:sq1}
\end{equation}

where \(O\), \(O_{sym}\), and \(O_{asym}\) are the sets of all orbits of \(\phi\), orbits that are displayed symmetrically in \(D\), and orbits that are displayed asymmetrically in \(D\) respectively; \(o_k,\), \(k = 1,2,...,|O|\) are the orbits of \(\phi\); and \(sd(o_k)\) denotes the Euclidean distance from the positions of points belonging to orbit \(o_k\) in \(D\) to exact symmetry. The second formula defines possible ranges of values for a drawing based on the number of orbits that are drawn as symmetric and is computed as:

\begin{equation}
SQ_2 = \begin{cases} 1 & \quad \text{if } | O_{sym} | = | O |\\ \frac{1 + |O_{sym}|}{|O|} - \left(1 - \sum_{o \in O_{asym}} sd(o)\right) & \quad \text{otherwise}\end{cases}
\label{eq:sq2}
\end{equation}

The fraction \(\frac{1 + |O_{sym}|}{|O|}\) limits the range of metric values based on the number of orbits displayed as symmetries in the drawing, while the sum \(\left(1 - \sum_{o \in O_{asym}} sd(o)\right)\) penalizes the results based on the Euclidean distance from exact symmetry.

\subsection{Symmetry Quality Metric for Automorphism Groups}

Another refinement of the metric is to take as input an \textit{automorphism group}, to detect the extent to which multiple automorphisms are simultaneously displayed in a drawing of a graph. Our metric is able to take as input dihedral groups, which contain both rotational and reflectional automorphisms.

To compute the group symmetry quality \(SQG\), we consider a weighted sum based on the maximum orbit size of each orbit set. Given \(A\) as the automorphism group to be checked, we define weight \(w\) as:

\begin{equation}
w = \sum_{\phi \in A} K(\phi)
\label{eq:weight}
\end{equation}

where \(K(\phi)\) is the size of the largest orbit of an automorphism \(\phi \in A\). For reflectional automorphisms, the value of \(K(\phi)\) is always 2, while for rotational automorphisms, the value of \(K(\phi)\) is the order of the rotation. Given this weight \(w\), the group symmetry quality score \(SQG\) is computed using the formula:

\begin{equation}
SQG = \begin{cases} \left(\frac{1}{w} \sum_{\phi \in A} K(\phi) \times SQ(\phi) \right) \times \frac{1}{2} & \quad \text{if } | A_{sym} | = 0\\ \left(1 + \frac{1}{w} \sum_{\phi \in A} K(\phi) \times SQ(\phi) \right) \times \frac{1}{2} & \quad \text{otherwise}\end{cases}
\label{eq:sqg}
\end{equation}
where \(A_{sym}\) is the subset of \(A\) containing automorphisms that are displayed as exact symmetries in the drawing \(D\) and \(SQ(\phi)\) is the score computed by the \(SQ\) metric with \(\phi\) as the input automorphism. \(SQ(\phi)\) can be computed with either \(SQ_1\) (Equation \ref{eq:sq1}) or \(SQ_2\) (Equation \ref{eq:sq2}).

The usage of the weight and the multiplier \(K(\phi)\) for each orbit is meant to give preference to drawings that display higher orders of symmetry, i.e. possessing more symmetries. Thus, a drawing that is symmetric but only of a lower order of symmetry will score lower than a symmetric drawing displaying a higher order of symmetry. Furthermore, adding 1 to the sum when at least an automorphism is realized as a symmetry ensures that every symmetric drawing will obtain a higher score than every asymmetric drawing.

\section{Algorithms for the Symmetry Quality Metric}
We present algorithms used to compute our metrics. Section \ref{sec:exactsymalg} and \ref{sec:appsymalg} provide algorithms for detecting exact and approximate symmetries respectively within the framework of our metric, Section \ref{sec:sqalg} presents the algorithm to compute the \(SQ\) version of our metric, and Section \ref{sec:sqgalg} presents the algorithm to compute the \(SQG\) version of our metric

\subsection{Exact Symmetry Detection}
\label{sec:exactsymalg}

While exact symmetry detection for two-dimensional point sets and polygons can be run in \(O(n \log n)\) time, node-link drawings of graphs are often not simple polygons, which necessitates a modification to the approach. With symmetric graph drawings, it is possible to detect an automorphism of the graph by detecting the geometric symmetry of the vertex point set of the drawing and checking the adjacency of the vertices. The following describes the algorithm:
\\
\\
\textbf{Algorithm 1: Algo-ExactSym}\\
\textbf{Input}: Graph \(G\), drawing \(D\) of \(G\)\\
\textbf{Output}: true or false
\begin{enumerate}
    \item Sort the vertex point set of \(D\) according to their angle around and distance from the centroid.
    \item \label{item:exactsymdet}Detect the geometric symmetry of the point set. If no symmetry is detected, return false, otherwise obtain the geometric symmetry \(\sigma\).
    \item \label{item:exactorbitcheck}For each orbit \(o_k\) in \(\sigma\), compare the adjacency of the vertices included in \(o_k\). If all of the vertices do not share the same pattern of adjacency, return false.
    \item Return true if for every orbit, all the vertices share the same adjacency pattern.
\end{enumerate}

\textit{Algo-ExactSym} runs in \(O(n \log n + m)\) time. Exact symmetry detection algorithms (both rotational and axial) for two-dimensional point sets, run in worst case \(O(n \log n)\) time~\cite{wolter1985optimal}. As each vertex only appears in one orbit, the total running time of the symmetry detection step is of \(O(n \log n)\) time in the number of vertices. Step \ref{item:exactorbitcheck} takes linear time in the degree of the vertices, i.e. linear in the number of edges, putting the time complexity of the algorithm as \(O(n \log n + m)\) time, with \(n\) as the number of vertices and \(m\) as the number of edges.

Recently, De Luca et al~\cite{deluca2019} presented a method to detect axial, rotational, and translational symmetries in graph drawings using a machine learning approach. However, this approach is a classification approach which only returns whether a drawing contains certain symmetries and with an 8\% misclassification rate.

\subsection{Approximate Symmetry Detection}
\label{sec:appsymalg}

The previous approach is limited by the nature of the exact symmetry detection algorithm, which either returns a positive answer for an exact symmetry or a negative answer otherwise and does not quantify how far a negative answer is to exact symmetry. There is therefore a need for a method that is able to compare the approximate symmetry of a drawing to the automorphism of the underlying graph.

Given an algorithm that computes the approximate symmetry of a point set, we can use it to determine whether each input orbit of automorphism is displayed as a geometric symmetry in the drawing of the graph. For our implementation, we use the symmetry distance approach of Zabrodsky et al~\cite{zabrodsky1995symmetry} and we compute \(sd\) as mentioned in the formulas of \(SQ_1\) and \(SQ_2\) as below:
\\
\\
\textbf{Algorithm 2: Algo-ApproxSym}\\
\textbf{Input}: Graph \(G\), drawing \(D\) of \(G\), automorphism \(\phi\) of \(G\)\\
\textbf{Output}: \(sd\)
\begin{enumerate}
    \item \label{item:normpoints}Normalize the points in the vertex point set \(P\) of drawing \(D\) such that all of the points lie within the unit circle centered on the centroid of the point set.
    \item For each orbit \(o_k\) in an automorphism \(\phi\):
    \begin{enumerate}
        \item Take the subset \(P_k\) of \(P\), containing points corresponding to vertices contained in \(o_k\).
        \item Compute the symmetric configuration \(P_\sigma\) closest to \(P_k\).
        \item For every vertex \(v\) included in \(o_k\), compute the Euclidean distance between its position in \(D\) and its image in \(P_\sigma\), average the values, and divide by 2 to obtain the average distance \(d\).
        \item Subtract \(d\) from 1 to obtain the value \(sd\).
    \end{enumerate}
\end{enumerate}

The distance computed by the symmetry distance approach is dependent on the area taken by the drawing \(D\). To normalize this distance, we scale the point set such that the whole drawing fits in a unit circle, which limits the maximum possible distance to 2. We then divide the average distance \(d\) by 2 and subtract the result from 1 in order to get a score where 1 corresponds to exact symmetry and lower values corresponds to drawings that are further from exact symmetry.

In theory, only orbits where the average distance \(d\) is 0 should be considered symmetric. However, computers work with floating point precision numbers rather than real numbers, leading to unavoidable round-off errors. To account for this, we define a threshold \(\epsilon\) such that orbits with \(d\) less than \(\epsilon\) are considered to be symmetric.

\textit{Algo-AproxSym} runs in \(O(n \log n)\). Given a center of rotation, the symmetry distance approach computes approximate rotational symmetry in \(O(n \log n)\) time due to the need to sort the points around the center. When the points are already sorted, the method takes \(O(n)\) time.

With axial symmetry, given an axis of symmetry and the orbits, the symmetry distance approach takes \(O(n)\) time to compute the symmetric image of each orbit and then compute the distance of each point to its image in the symmetric configuration. 

\subsection{Algorithm for Computing \(SQ\)}
\label{sec:sqalg}

We present an algorithm to compute the \(SQ\) metrics for symmetry detection with a single automorphism as the input. We take a single set of orbits denoting one automorphism of a graph and compute how faithfully a drawing of the graph depicts this particular automorphism as a symmetry. Given a drawing \(D\) of a graph \(G\) and an automorphism \(\phi\) of \(G\), the metric is computed as follows:
\\
\\
\textbf{Algorithm 3: Algo-SQ}\\
\textbf{Input}: Graph \(G\), drawing \(D\) of \(G\), automorphism \(\phi\) of \(G\)\\
\textbf{Output}: \(SQ\) metric score
\begin{enumerate}
    \item For each orbit \(o_k\) in \(\phi\):
    \begin{enumerate}
        \item Compute the symmetry distance-based score \(sd\) using \(Algo-ApproxSym\)
        \item Add \(o_k\) to the set \(O_{sym}\) if the value of \(sd\) is less than or equal to a threshold \(\epsilon\), or add it to the set \(O_{asym}\) otherwise
    \end{enumerate}
    \item Compute the symmetry quality using either\(SQ_1\) or \(SQ_2\).
\end{enumerate}

Using \(Algo-ApproxSym\), given a center or axis of symmetry, the \(sd\) values can be computed in \(O(n \log n)\) time in the number of vertices. When this information is not given, we select the center or axis of symmetry in the following way:
\begin{itemize}
  \item For rotational symmetries, we compute the centroids for each orbit (the point itself in the case of fixed points) and select one that is the closest to all other centres (i.e. geometric median). In the case of ties, we compute the \(sd\) of each orbit and select the one with the largest \(sd\) score. After initially sorting the points in \(O(n \log n)\) time, computing the geometric median can be done with two \(O(n)\)-time sweeps along the x- and y-axes and computing the \(sd\) scores when needed takes \(O(n)\) time , keeping the runtime at \(O(n \log n)\) time.
  \item For axial symmetries, we compute the best line of symmetry using Singular Value Decomposition (SVD) as a pre-processing step, as proposed by Zabrodsky et al~\cite{zabrodsky1995continuous}. While SVD runs in \(O(n^3)\) in the size of the input matrices, in our case the matrix is of a fixed size \(2 \times 2\) as we only consider drawings in 2 dimensions, keeping the runtime complexity of the metric computation at \(O(n \log n)\) time in the number of vertices.

Therefore, \textit{Algo-SQ} runs in \(O(n \log n)\) time.
\end{itemize}

\subsection{Algorithm for Computing \(SQG\)}
\label{sec:sqgalg}

The \(SQG\) metric takes as input multiple automorphisms of a graph that can be displayed simultaneously in the same drawing and returns a symmetry quality metric score that measures how faithfully the drawing displays all the automorphisms at once. Given a drawing \(D\) of a graph \(G\) and a group of automorphisms \(A\) of \(G\), the metric is computed using the following steps:
\\
\\
\textbf{Algorithm 4: Algo-SQG}\\
\textbf{Input}: Graph \(G\), drawing \(D\) of \(G\), group of automorphisms \(A\) of \(G\)\\
\textbf{Output}: \(SQG\) metric score
\begin{enumerate}
    \item For each automorphism \(\phi\) in \(A\), compute \(SQ(\phi)\) using \(Algo-SQ\).
    \item Compute the weighted sum of \(SQ(\phi)\) for all automorphisms in \(A\), where the scores are weighted by the maximum orbit size of each automorphism.
    \item Compute the \(SQG\) using Equation \ref{eq:sqg}.
\end{enumerate}

The computation of each \(SQ\) score for each automorphism runs in \(O(n \log n)\) time. With \(k\) as the number of automorphisms in the input group, the \(SQG\) computation as a whole runs in \(O(kn \log n)\) time.

\section{Experiment 1: \(SQ\) Metric Validation Experiments}
\label{sec:singlefixedvalidation}
\subsection{Experiment Design}
To validate the \(SQ\) version of our metric, we perform experiments where we take a symmetric drawing of a graph and deform it in a way that breaks the symmetry.

We expect that not only will our metric effectively capture the distortion from perfect symmetry induced by the deformations, but also that it will perform better than existing approximate symmetry detection approaches which only rely on Euclidean distance, here represented as \(sd\). We formulate the following hypotheses:

\begin{itemize}
  \item Hypothesis 1: The symmetry quality scores \(SQ_1\) and \(SQ_2\) will both decrease as the drawing \(D\) of \(G\) is further deformed.
  \item Hypothesis 2: The symmetry quality scores \(SQ_1\) and \(SQ_2\) will reflect the extent of distortion from perfect symmetry more effectively than \(sd\).
\end{itemize}

We performed five sets of experiments each to validate our metric on rotational and axial symmetry. For each experiment, we perform the following steps:
\begin{enumerate}
\item For a symmetric graph \(G\), select \(\phi\), one of its geometric automorphisms.
\item Create a drawing \(D\) of \(G\) that displays \(\phi\) as an exact symmetry.
\item Select a subset of vertices of \(G\) and perturb their position by taking them further from their initial position. Repeat until a desired number of perturbation steps is reached.
\end{enumerate}

We used a mix of well-known symmetric graphs from graph theory literature and newly-generated symmetric graphs. The graphs we generated for validation experiments with rotational symmetry (titled in the format \(c[order]x[\# of orbits]\)) were generated as follows. First, we decide on the order of rotational symmetry \(k\) and the number of orbits \(m\), then define a graph \(G\) with \(k \times m\) vertices and initially zero edges. We then connect each of the \(m\) sets of \(k\) vertices into \(m\) distinct cycles of length \(k\), and, except for the ``innermost'' cycle, we connect them to previous cycles with a number of edges between \(k\) and \(2k\) such that the graph possesses the order of symmetry desired.

With graphs generated for axial symmetry, we first determine the orbits, including fixed points, create edges between vertices not in the same orbit, then ``mirror'' the edges by adding edges between the points which share orbits with the endpoint of the created edges.

\subsection{Rotational Symmetry Experiments}

\begin{figure}[H]
    \centering
    \begin{tabular}{|c|c|}
    \hline
     Step 0 & Step 3 \\ \hline
     \includegraphics[width=0.4\columnwidth]{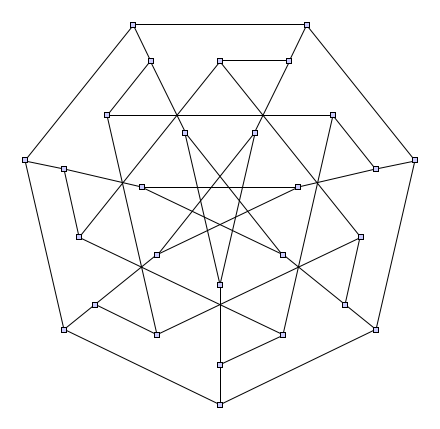} & \includegraphics[width=0.4\columnwidth]{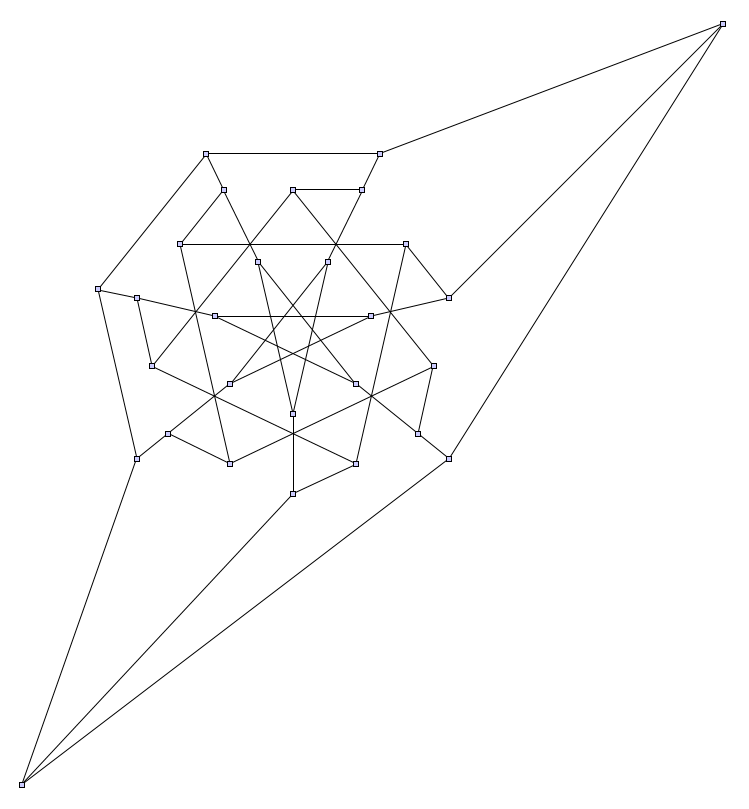}\\ \hline
     Step 7 & Step 10 \\ \hline
     \includegraphics[width=0.4\columnwidth]{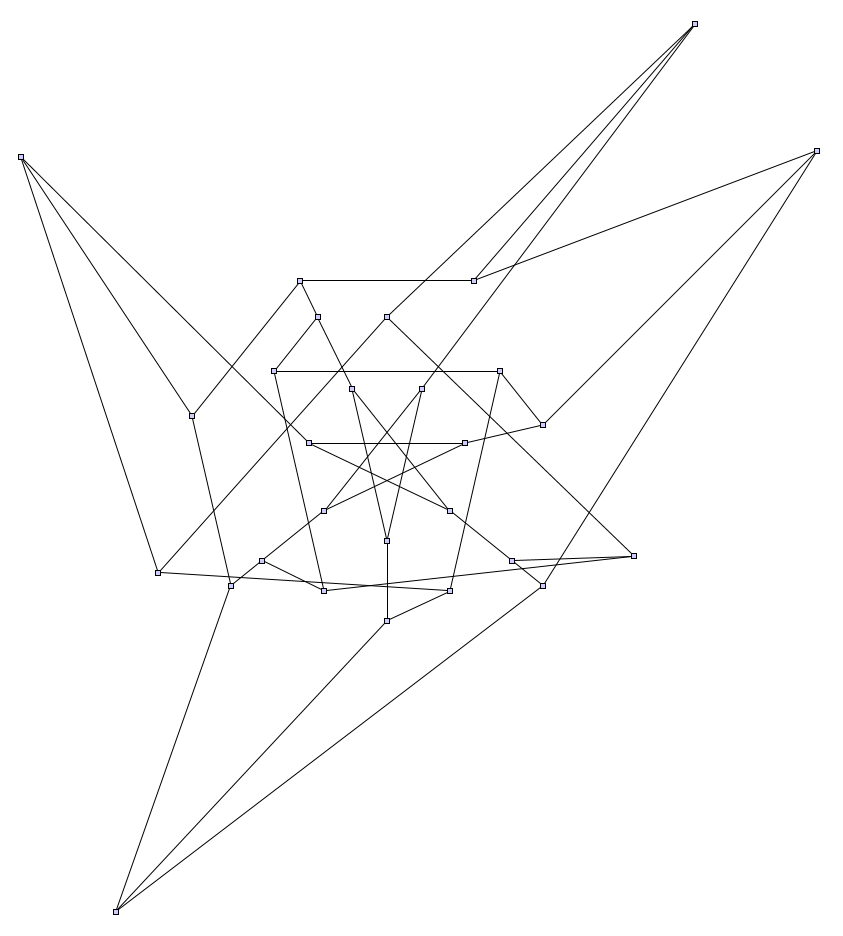} & \includegraphics[width=0.4\columnwidth]{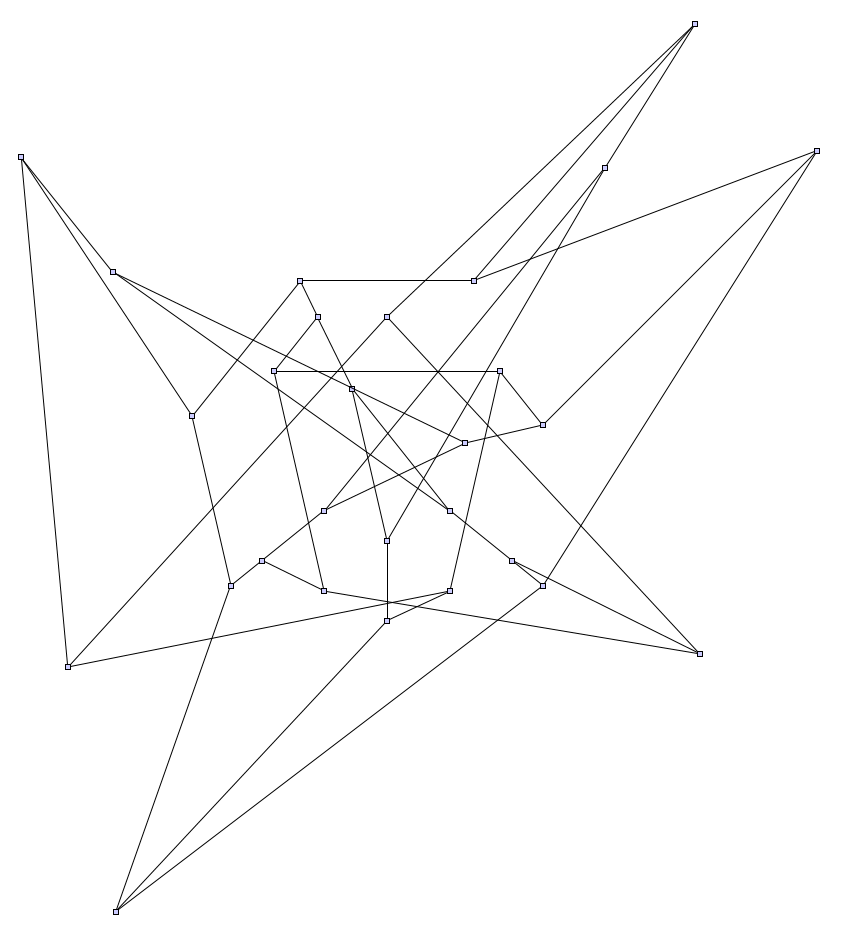} \\ \hline
    \end{tabular}
    \caption{Validation experiments detecting rotational symmetry (order 7) on Coxeter graph. Each subsequent perturbation destroys a new orbit or further perturbs an already destroyed one.}
    \label{fig:coxeter_c7_ver}
\end{figure}

\begin{figure}[H]
    \centering
    \includegraphics[width=0.9\columnwidth]{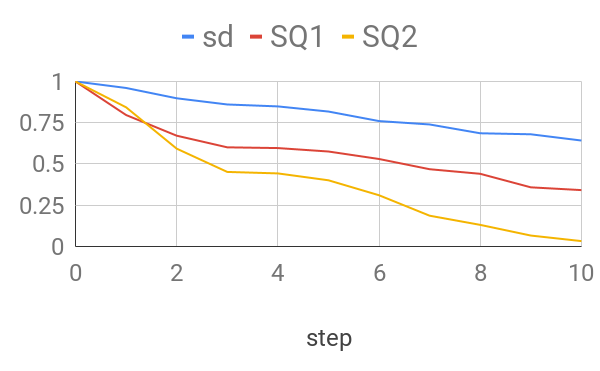}
    \caption{Symmetry quality metrics for the experiment in Figure \ref{fig:coxeter_c7_ver}. Both \(SQ_1\) and \(SQ_2\) captures the effect of the perturbations better than \(sd\) which is only based on Euclidean distance, with \(SQ_2\) capturing the effects of further perturbations better.}
    \label{fig:coxeter_c7_ver_scores}
\end{figure}

Figure \ref{fig:coxeter_c7_ver} shows an example of the validation experiment for detecting rotational symmetries. In this example, we take the Coxeter graph~\cite{coxeter1983my}, which possesses an automorphism that can be depicted as a rotational symmetry of order 7 with 4 orbits. Step 0 shows the initial drawing of the graph, where all four orbits of order 7 rotation are displayed in the drawing. In step 3, one orbit has been perturbed, in step 7, two more orbits have been perturbed, and in step 10, all orbits have had two each of their vertices perturbed.

Figure \ref{fig:coxeter_c7_ver_scores} shows the plot of the \(SQ\) metrics computed for the experiment displayed in Figure \ref{fig:coxeter_c7_ver}. We plot the \(SQ\) metrics computed by both of our formulas, \(SQ_1\) and \(SQ_2\) together with the \(sd\) score computed solely using the Euclidean distance as comparison.

It can be seen that not only both \(SQ_1\) and \(SQ_2\) scores decrease as the perturbation steps progress, supporting Hypothesis 1, they also decrease at a faster rate than \(sd\). While \(sd\) only decreases to around 0.6 at the final step, \(SQ_1\) decreases to around 0.3 and \(SQ_2\) to less than 0.1. These scores more accurately reflects the state of the drawing at Step 10 in Figure \ref{fig:coxeter_c7_ver}, which is visibly far from perfect symmetry, and thus supporting Hypothesis 2.

\subsection{Axial Symmetry Experiments}
\begin{figure}[ht]
    \centering
    \begin{tabular}{|c|c|}
    \hline
     Step 0 & Step 3 \\ \hline
     \includegraphics[height=0.35\columnwidth]{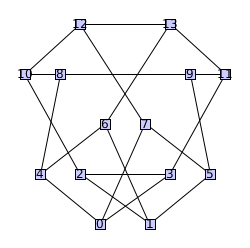} & \includegraphics[height=0.35\columnwidth]{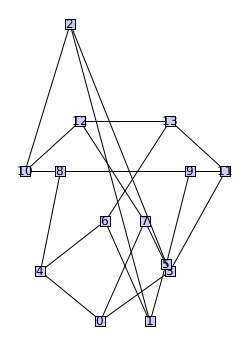}\\ \hline
     Step 7 & Step 10 \\ \hline
     \includegraphics[height=0.35\columnwidth]{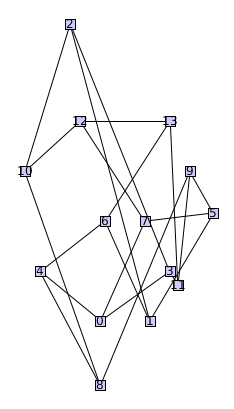} & \includegraphics[height=0.35\columnwidth]{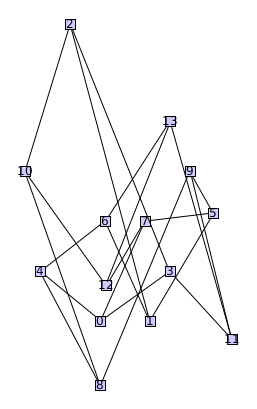} \\ \hline
    \end{tabular}
    \caption{Validation experiment detecting axial symmetry on Heawood graph.}
    \label{fig:heawood_ver}
\end{figure}
\begin{figure}[ht]
    \centering
    \includegraphics[width=0.9\columnwidth]{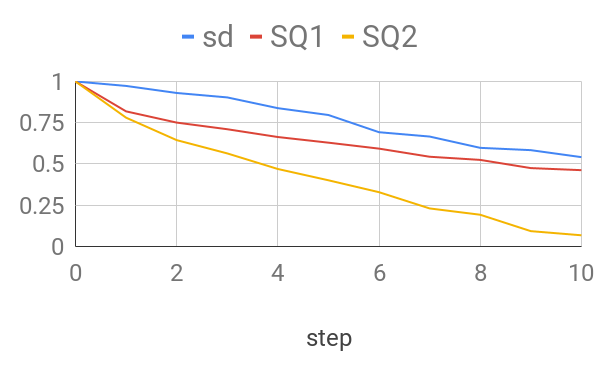}
    \caption{Symmetry quality scores for the validation experiment detecting axial symmetry on Heawood graph. \(SQ_2\) decrease at a faster rate on latter perturbation steps than \(SQ_1\)}
    \label{fig:heawood_ver_scores}
\end{figure}

Figure \ref{fig:heawood_ver} shows an example of a validation experiment for using the \(SQ\) metric with an axial automorphism as input. We use the Heawood graph in this example and draw a layout that displays one of its automorphisms as an axial symmetry, with the y-axis as the axis of symmetry. Each subsequent drawing in Figure \ref{fig:heawood_ver} destroys at least one more orbit compared to the previous drawing.

Figure \ref{fig:heawood_ver_scores} shows the \(SQ\) metrics computed for each perturbation step for the experiment depicted in Figure \ref{fig:heawood_ver}. Similar to the case seen in Figure \ref{fig:coxeter_c7_ver_scores}, both \(SQ_1\) and \(SQ_2\) decrease as the perturbation steps continue and both obtain lower scores than \(sd\) on the perturbed drawings, further supporting Hypotheses 1 and 2. Similar patterns can be seen in all other experiments, validating Hypotheses 1 and 2.

\subsection{Discussion and Summary}

In all validation experiments, \(SQ_1\) and \(SQ_2\) decrease steadily with the number of perturbation steps, while also capturing the changes in quality better than \(sd\). Another point of note is that \(SQ_2\) consistently decreases at a faster pace than \(SQ_1\), such as can be seen in Figure \ref{fig:coxeter_c7_ver_scores}, denoting that it can be more effective at capturing the changes in the quality of the symmetry of the drawings.

Furthermore, in cases such as in Figure \ref{fig:heawood_ver_scores}, we see that on the later half of the perturbation steps, the scores computed by the \(SQ_1\) formula does not differ much from \(sd\), while \(SQ_2\) continues to decrease to a low score of below 0.1 for step 10 where five out of seven orbits have been perturbed. This further shows that \(SQ_2\) is better at capturing distortions from perfect symmetry.

\textit{In summary, our experiments have supported Hypotheses 1 and 2, validating the effectiveness of our metric to measure the quality of symmetry in graph drawings with respect to one selected graph automorphism and demonstrates its better effectiveness in measuring deviations from perfect symmetry compared to the Euclidean distance-based-only \(sd\), as well as showing that \(SQ_2\) captures the changes in quality more effectively than \(SQ_1\).}

\section{Experiment 2: \(SQG\) Metric Validation Experiments}

\subsection{Experiment Design}

\begin{figure*}[ht]
    \centering
    \begin{tabular}{|c|c|c|c|c|}
    \hline
     \(D_1\) & \(D_2\) &
     \(D_3\) & \(D_4\) & \(D_5\)  \\ \hline
     \includegraphics[height=0.175\textwidth]{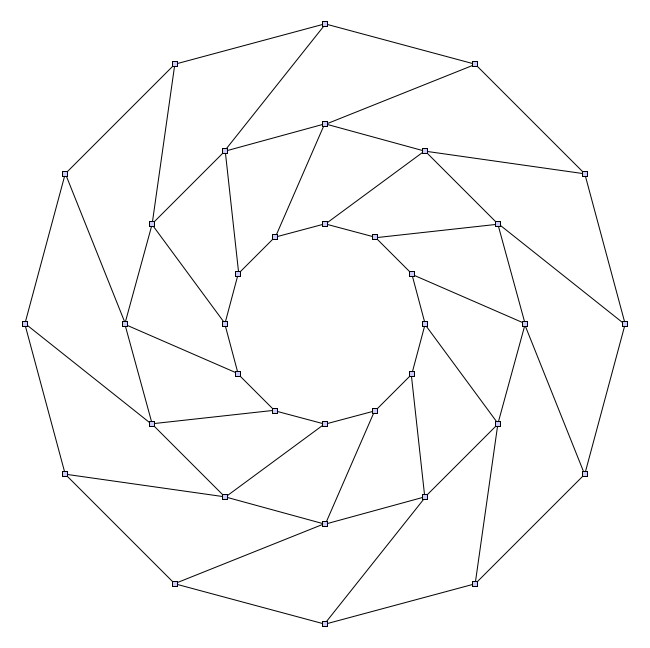} & \includegraphics[height=0.175\textwidth]{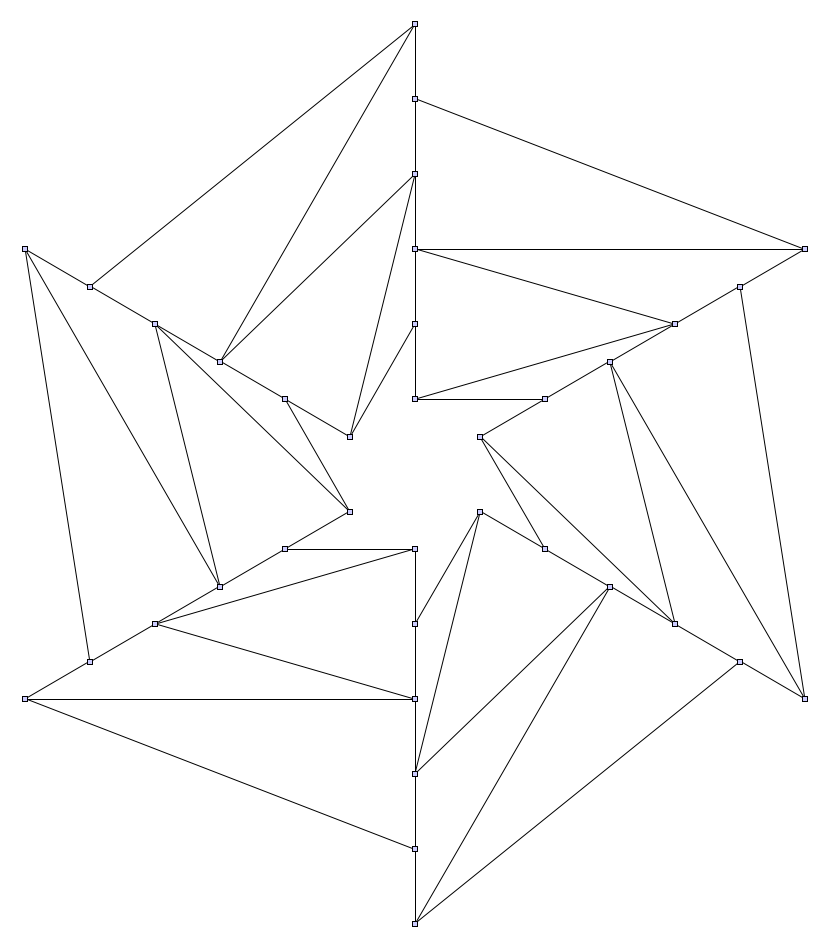} &
     \includegraphics[height=0.175\textwidth]{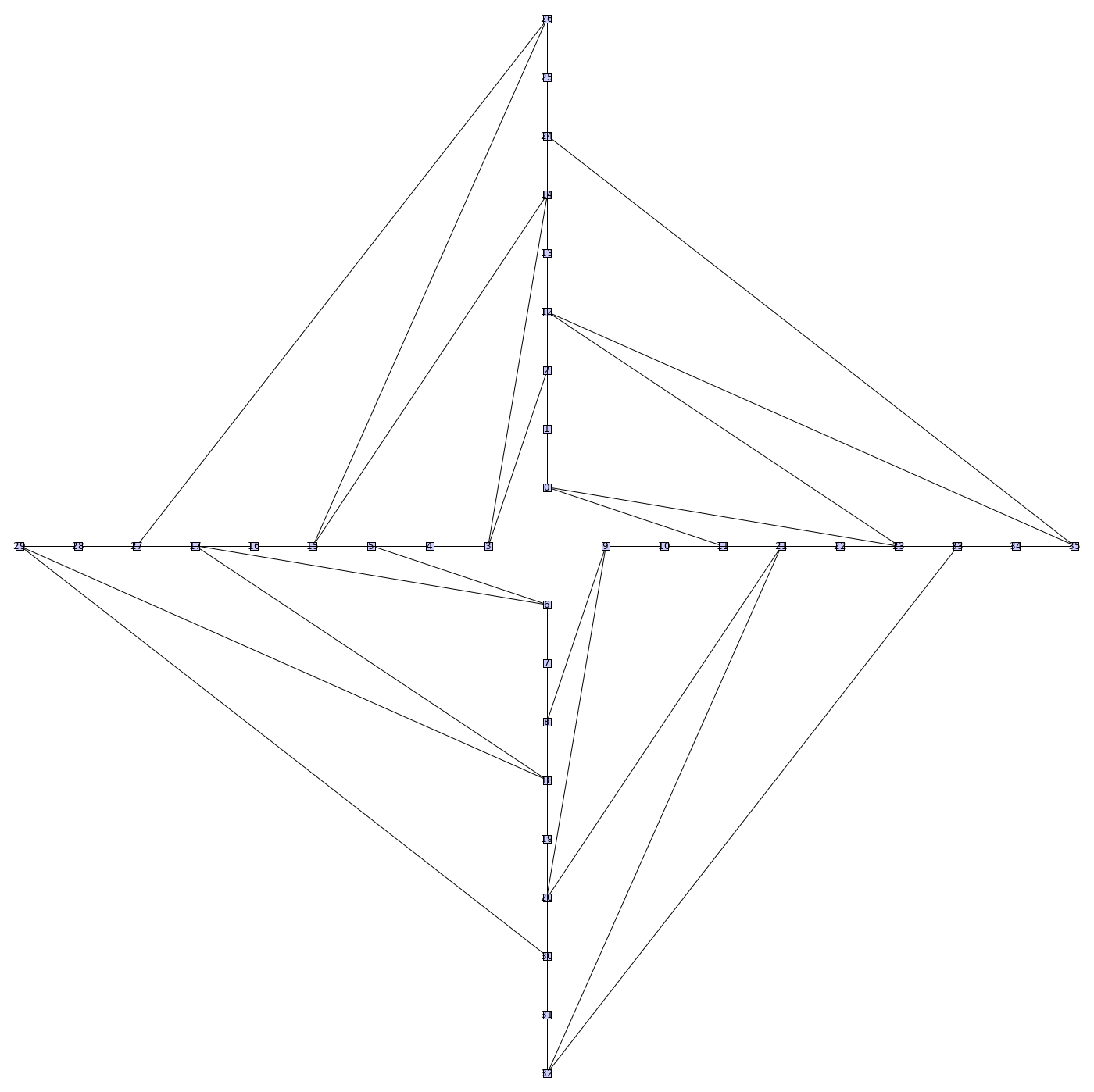} & \includegraphics[height=0.175\textwidth]{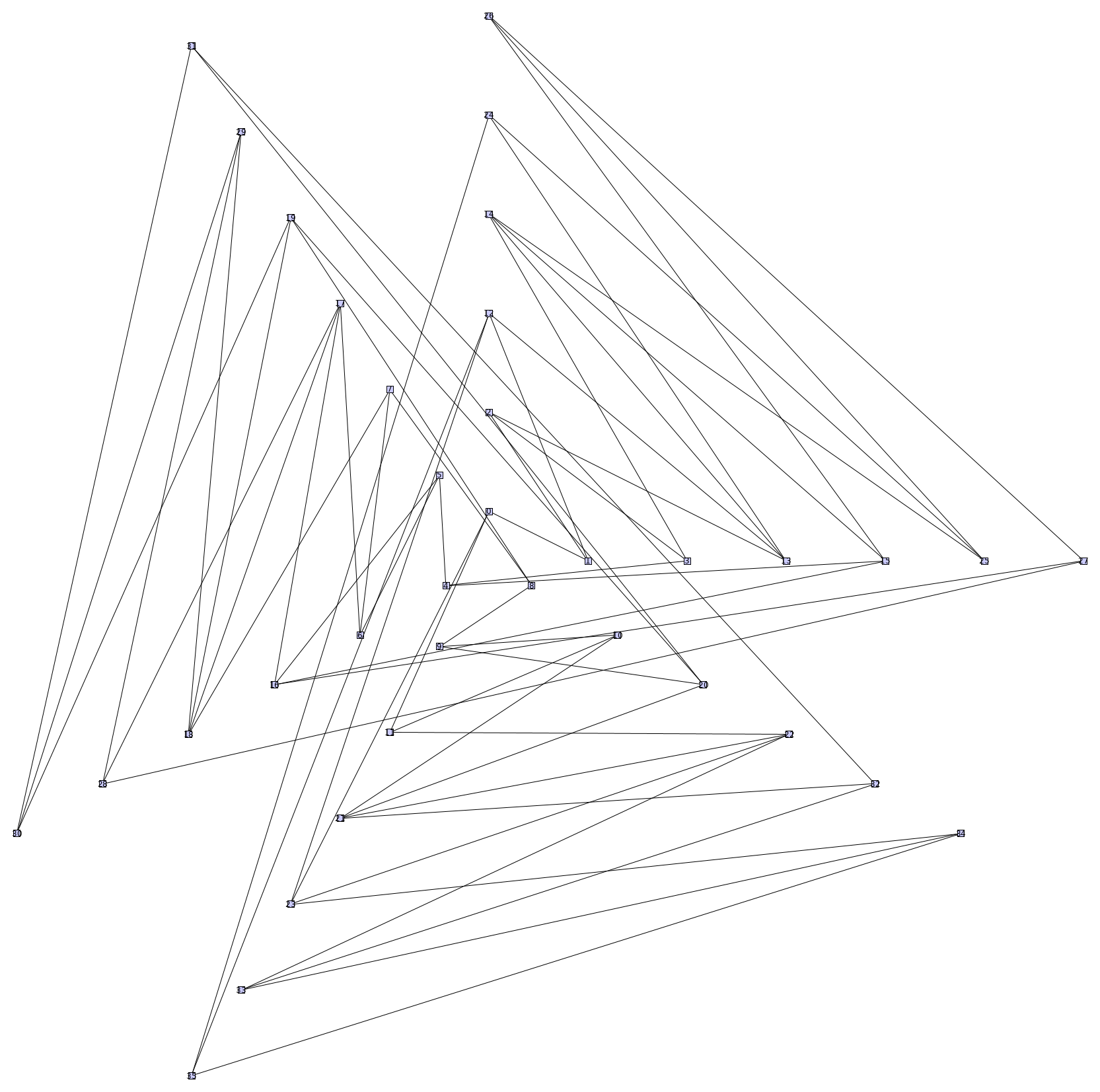} &
     \includegraphics[height=0.175\textwidth]{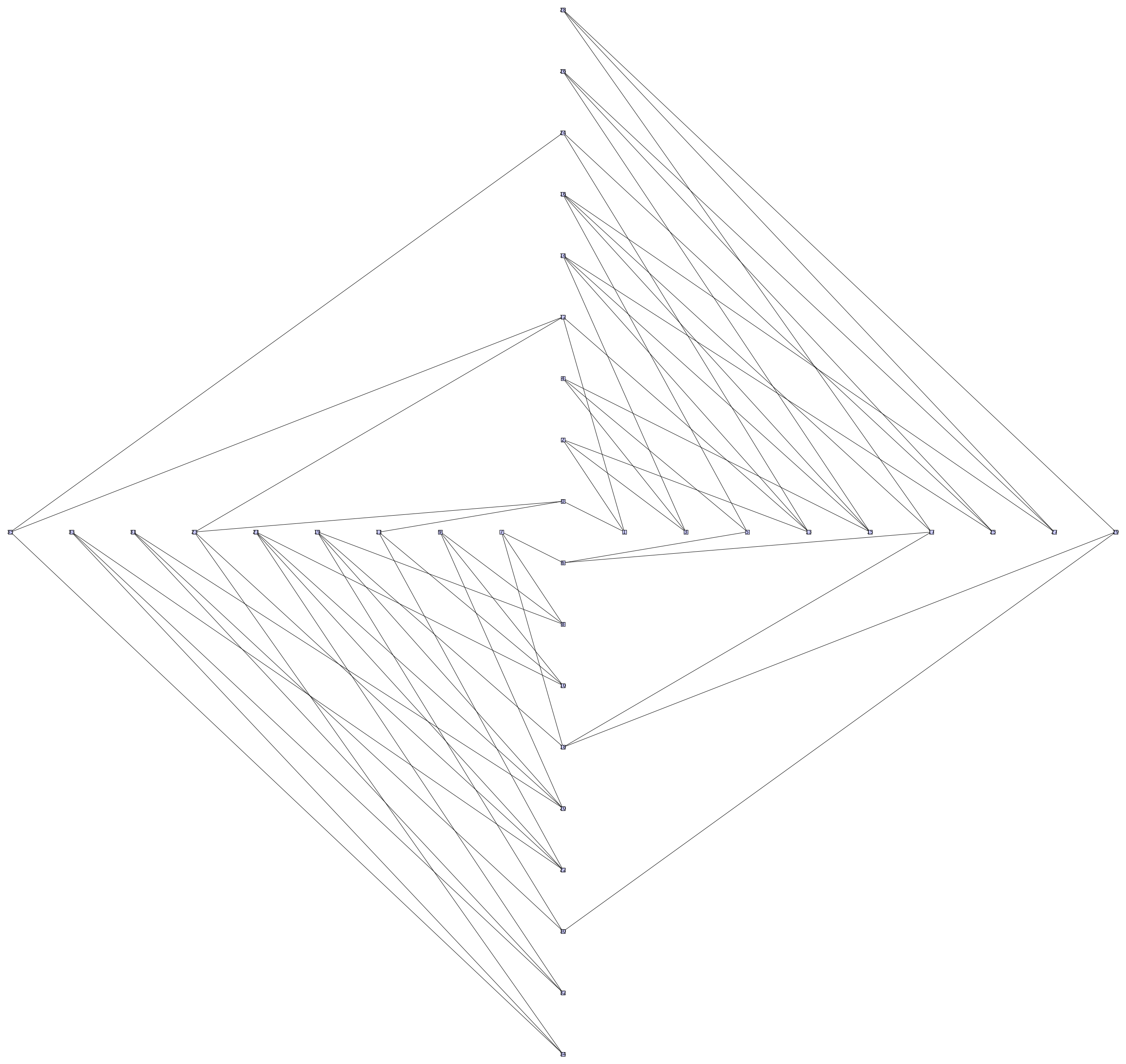} \\ \hline
    \end{tabular}
    \caption{Validation experiment with rotational symmetries of orders 12, 6, 4, 3, and 2.}
    \label{fig:c12x3_subrot}
\end{figure*}

\begin{figure*}[ht]
    \centering
    \begin{tabular}{|c|c|c||c|c|c|}
    \hline
    \multicolumn{3}{c}{Dodecahedral graph} & \multicolumn{3}{c}{Cuboctahedral graph} \\ \hline
     \(D_1\) & \(D_2\) & \(D_3\) & \(D_1\) & \(D_2\) & \(D_3\) \\ \hline
     \includegraphics[width=0.3\columnwidth]{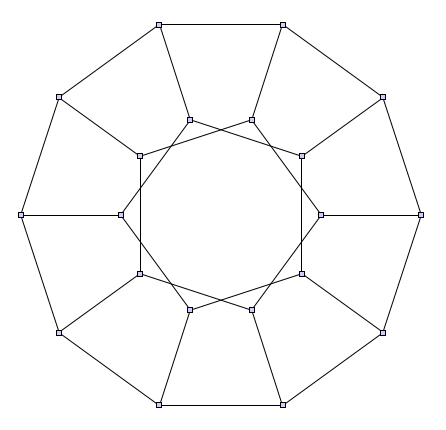} &
     \includegraphics[width=0.3\columnwidth]{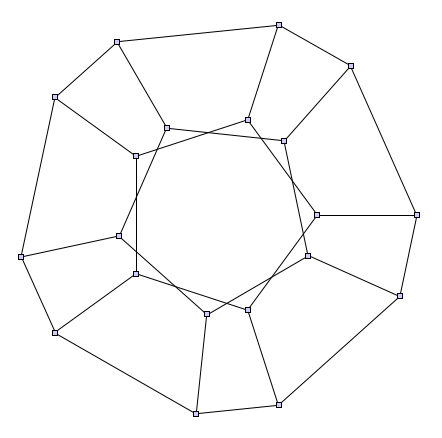} &
     \includegraphics[width=0.3\columnwidth]{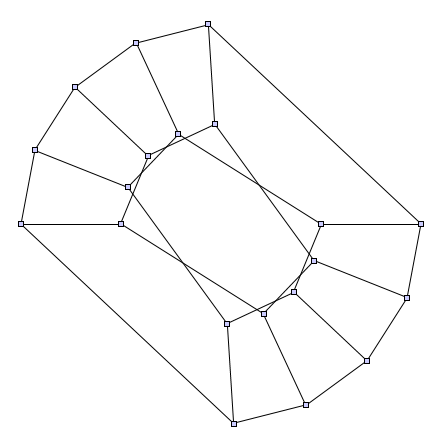} & \includegraphics[width=0.3\columnwidth]{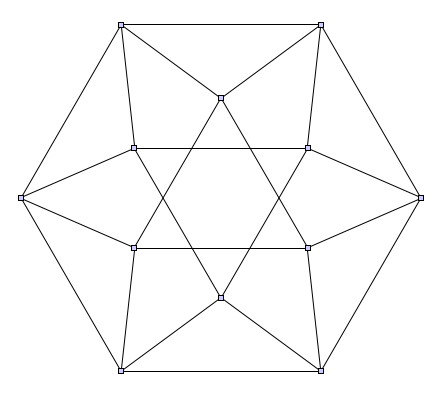} & \includegraphics[width=0.3\columnwidth]{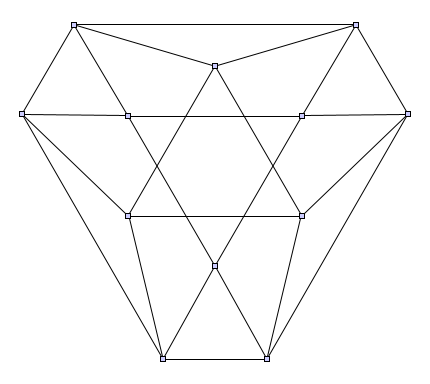} & \includegraphics[width=0.3\columnwidth]{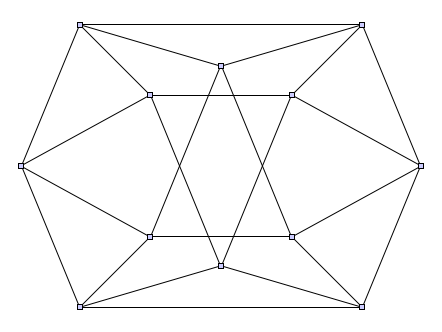} \\ \hline
    \end{tabular}
    \caption{Validation experiments with dihedral symmetry of orders 10, 5, and 2 on the dodecahedral graph and dihedral symmetry groups of orders 6, 3, and 2 on the cuboctahedral graph.}
    \label{fig:dodecahedral_cuboctahedral_subrot}
\end{figure*}

\begin{figure*}[ht]
    \centering
    \begin{tabular}{|c|c|c||c|c|c|}
    \hline
    \multicolumn{3}{c}{c12x3 (rotational order 12)} & \multicolumn{3}{c}{c12x3 (rotational order 6)} \\ \hline
     \(C12D_1\) & \(C12D_2\) & \(C12D_3\) & \(C6D_1\) & \(C6D_2\) & \(C6D_3\) \\ \hline
     \includegraphics[width=0.3\columnwidth]{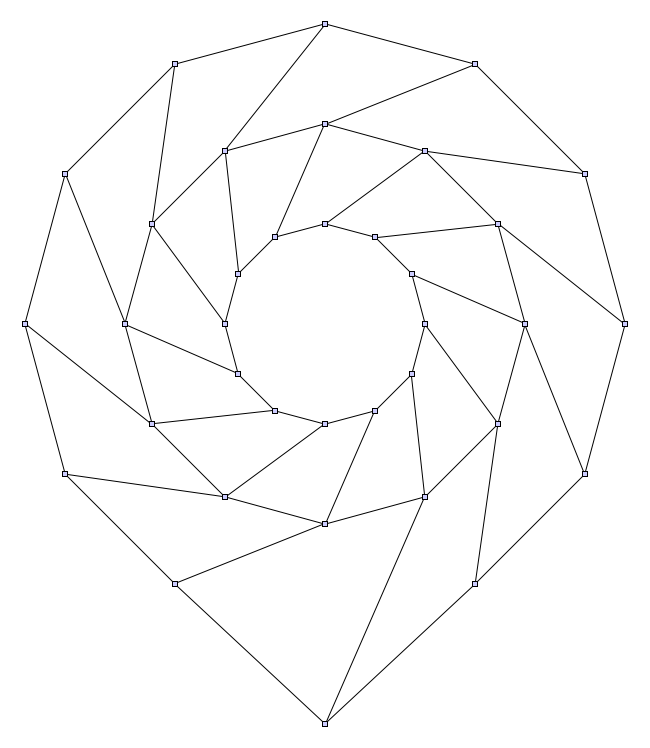} &
     \includegraphics[width=0.3\columnwidth]{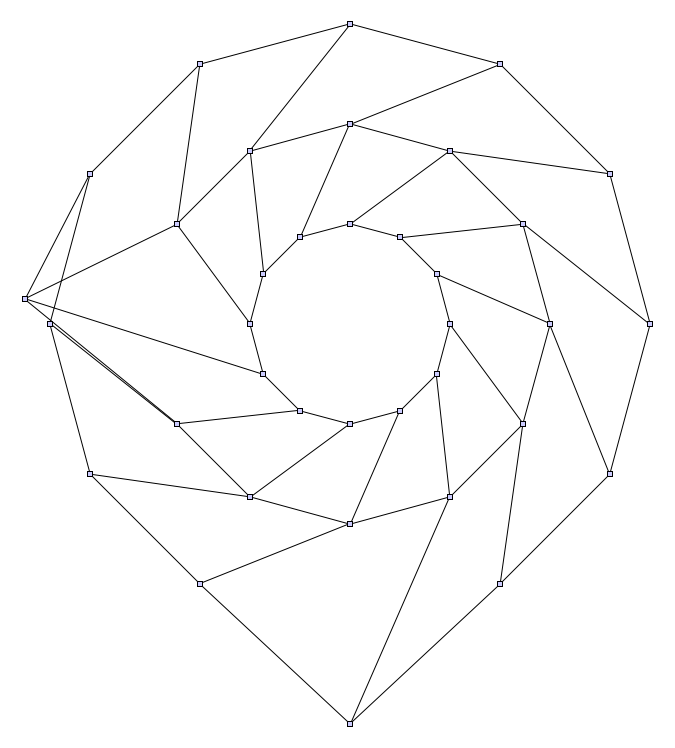} &
     \includegraphics[width=0.3\columnwidth]{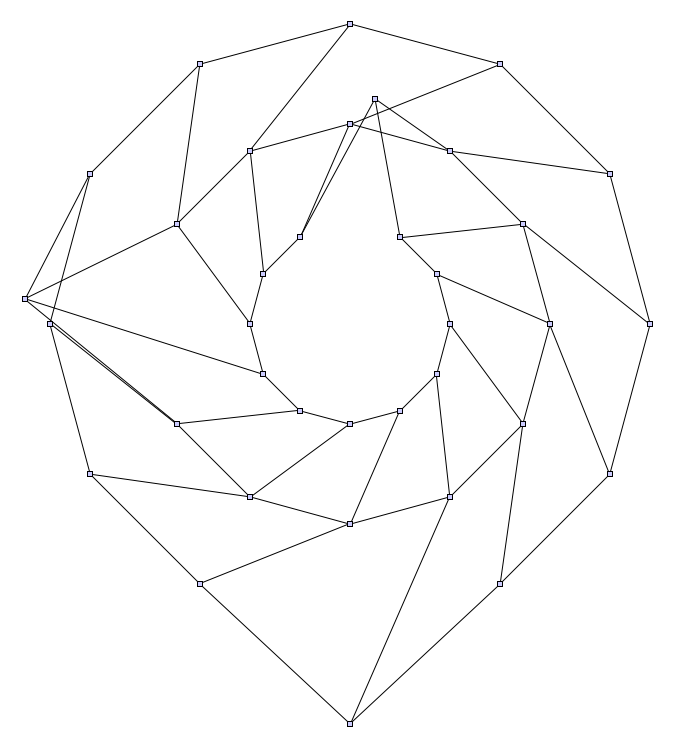} & \includegraphics[width=0.3\columnwidth]{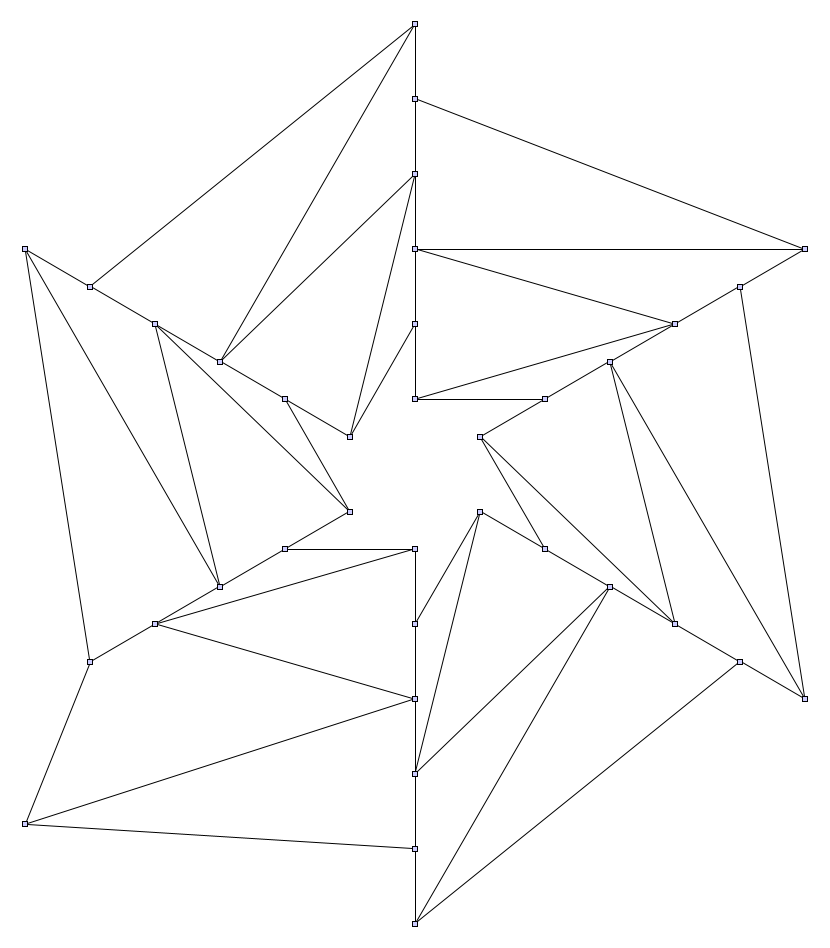} & \includegraphics[width=0.3\columnwidth]{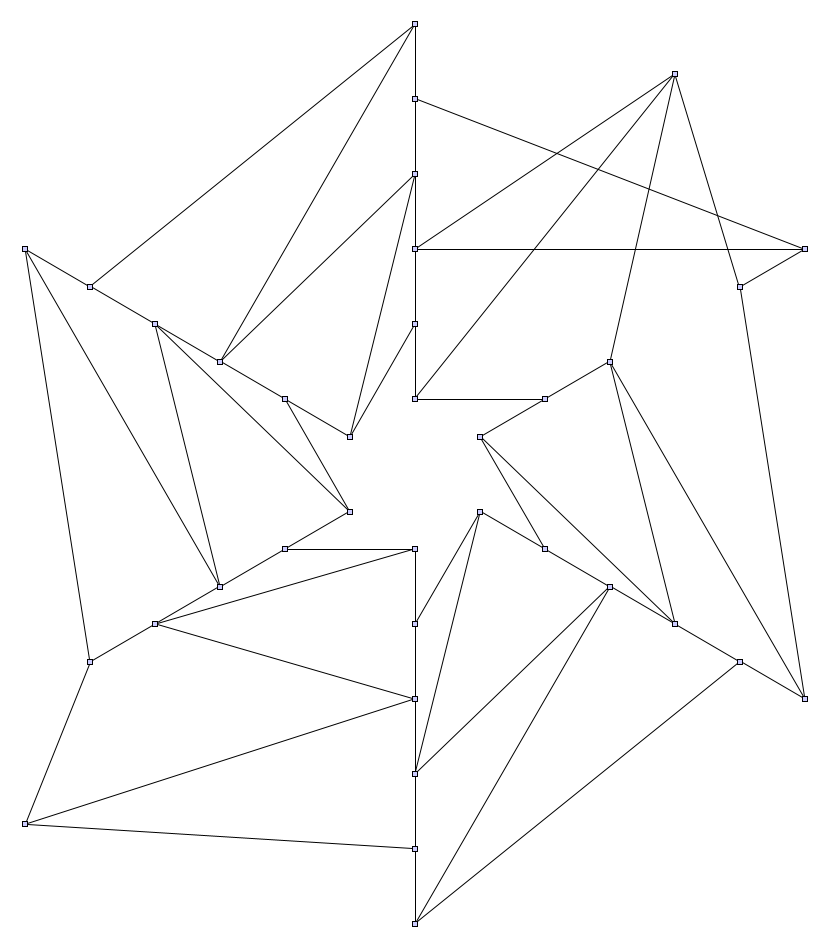} & \includegraphics[width=0.3\columnwidth]{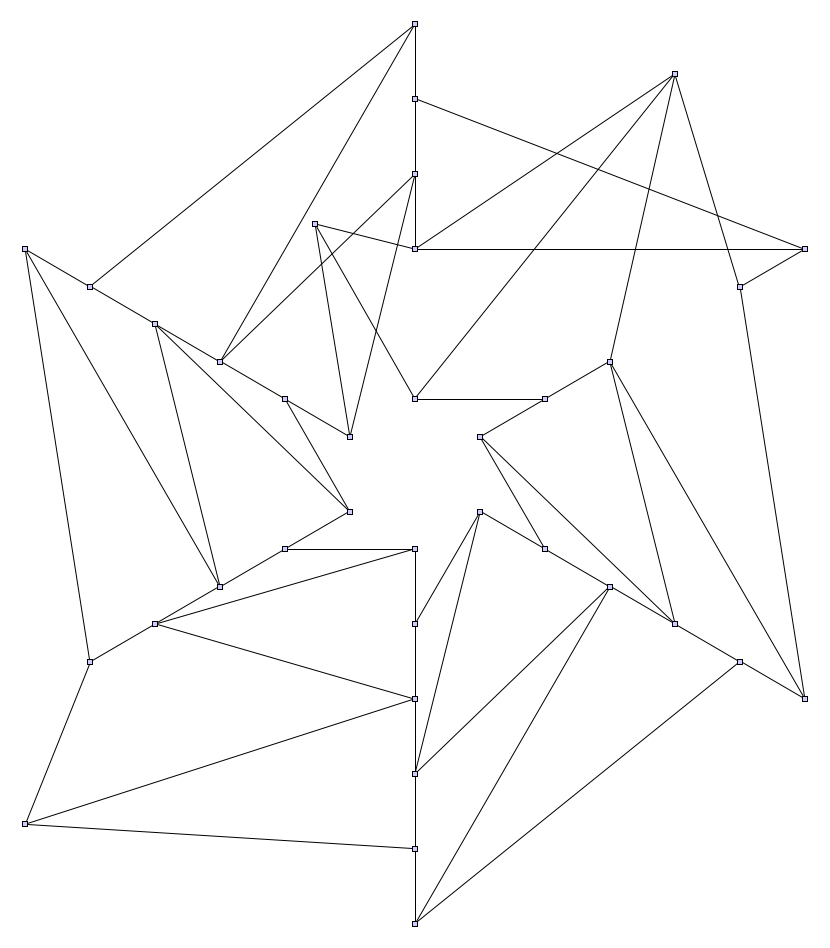}\\ \hline
    \end{tabular}
    \caption{Validation experiments with rotational symmetry of orders 12, 6, 4, 3, and 2 on a synthetic symmetric graph. Here, we start with drawings as defined in Figure \ref{fig:c12x3_subrot} and destroy another orbit of symmetry in each steps.}
    \label{fig:c12x3_subrotdist}
\end{figure*}

\begin{figure*}[ht]
    \centering
    \subfloat[c12x3 (rotational symmetry group detection)]{
        \includegraphics[width=0.5\columnwidth]{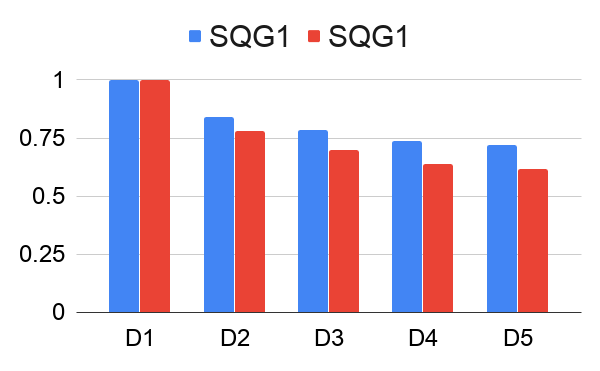}
    }
    \subfloat[dodecahedral graph]{
        \includegraphics[width=0.5\columnwidth]{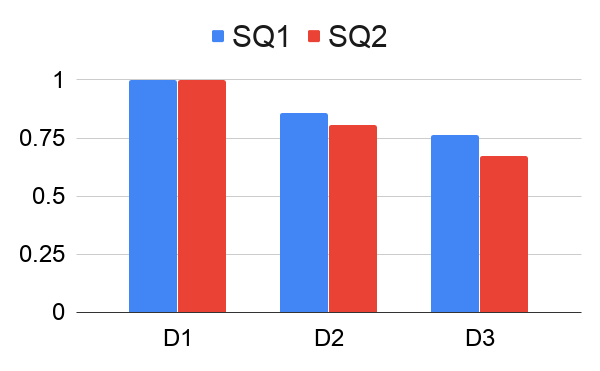}
    }
    \subfloat[cuboctahedral graph]{
        \includegraphics[width=0.5\columnwidth]{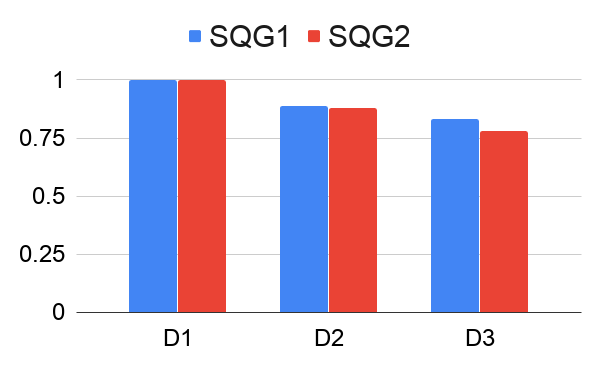}
    }
    \subfloat[c12x3 (with perturbation)]{
        \includegraphics[width=0.5\columnwidth]{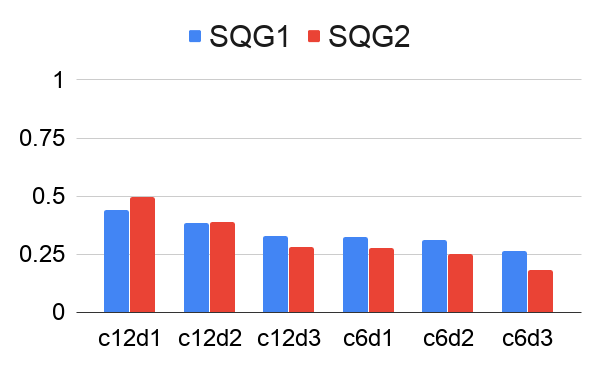}
    }
    \caption{Symmetry quality metrics computed for (a) c12x3 (Fig. \ref{fig:c12x3_subrot}, (b) dodecahedral graph (Fig. \ref{fig:dodecahedral_cuboctahedral_subrot}, (c) cuboctahedral graph (Fig. \ref{fig:dodecahedral_cuboctahedral_subrot}, (d) c12x3 (with perturbations) (Fig. \ref{fig:c12x3_subrotdist}}
    \label{fig:symgroup_val_scores}
\end{figure*}

To validate the \(SQG\) version of our metric, we compare drawings that show different subsets of an automorphism group and compare the value of the metric computed on each drawing.

We expect that our metric will score a drawing higher the more automorphisms from the input automorphism group it displays as symmetries, or if it displays an automorphism with a higher order. Furthermore, as with the \(SQ\) version, we expect our metric to effectively reflect the degree of approximate symmetry of a drawing and score asymmetric drawings that are closer to exact symmetry higher. We formulate the following hypotheses:

\begin{itemize}
  \item Hypothesis 3: The symmetry quality scores \(SQG_1\) and \(SQG_2\) will be higher for a drawing \(D_i\) that displays as symmetries a larger subset of the automorphism group or an automorphism with a larger order than a drawing \(D_j\) displaying as symmetries a smaller subset of the automorphism group or an automorphism with a smaller order.
  \item Hypothesis 4: The symmetry quality scores \(SQG_1\) and \(SQG_2\) will decrease as a drawing \(D\) is further deformed.
\end{itemize}

For these experiments, we start by selecting a symmetric graph \(G\) and a geometric automorphism group \(A\). We then create drawings of \(G\) displaying different subsets of \(A\) as symmetries. In addition, we also perform deformation experiments similar to that described in Section \ref{sec:singlefixedvalidation}, where we perform steps of deformation that gradually destroys more orbits and/or brings perturbed vertices further from their initial positions. We perform six sets of experiments with drawings displaying the input automorphism groups as exact symmetries and four sets of experiments with drawing deformations.

\subsection{Cyclic Group Experiments}

Figure \ref{fig:c12x3_subrot} shows an example of the validation experiment for the automorphism group testing version of the metric, using a graph we created called \(c12x3\) with a rotational automorphism of order 12 and 3 orbits. In this experiment, the input automorphisms are cyclic groups of order 12, 6, 4, 3, 2. In accordance, we created a set of drawings where the largest order of rotational symmetry displayed in each drawing correspond to one of the cyclic group orders in the input.

Figure \ref{fig:symgroup_val_scores}(a) shows the \(SQG\) metrics computed for the experiment displayed in Figure \ref{fig:c12x3_subrot}. The largest order of rotational symmetry displayed by each drawing is lower than the previous drawing - the first drawing, \(D_1\), displays rotational symmetry of order 12, while the last drawing, \(D_5\), only displays rotational symmetry of order 2. It can be seen that both \(SQG_1\) and \(SQG_2\) computes lower scores for drawing where the largest order of symmetry displayed is smaller, in line with Hypothesis 3.

\subsection{Dihedral Group Experiments}

Figure \ref{fig:dodecahedral_cuboctahedral_subrot} displays two more sets of validation experiments, where the input automorphisms are dihedral groups of order 10, 5, and 2 on the dodecahedral graph and dihedral groups of order 6, 3, and 2 on the cuboctahedral graph. In both examples, we start with a drawing displaying the highest order of symmetry by drawing each orbit on concentric circles. For subsequent drawings, we rotate a number of vertices in each orbit such that the drawing only shows a lower order of symmetry.

Figures \ref{fig:symgroup_val_scores}(b) and (c) shows the \(SQG\) scores computed for the validation experiments in Figure \ref{fig:dodecahedral_cuboctahedral_subrot}. It can be seen that for both sets of experiments, the scores computed by both \(SQG_1\) and \(SQG_2\) again decrease with drawings displaying lower orders of symmetry while still staying above 0.5, supporting Hypothesis 3.

\subsection{Automorphism Group with Perturbations Experiments}

Figure \ref{fig:c12x3_subrotdist} show an example of one automorphism group detection validation experiment with perturbations of the drawing. We start with drawings shown in Figure \ref{fig:c12x3_subrot}, in this case taking \(D_1\) and \(D_2\), then perform deformation steps by selecting an orbit, selecting a vertex from the orbit, and randomly perturbing its position such that the orbit is no longer displayed as symmetric in the drawing. Here, we show three steps of perturbation for each drawing.

Figure \ref{fig:symgroup_val_scores}(d) shows the \(SQG\) metrics computed on the drawings. It can be seen that for each set, the computed values gradually decrease as more orbits are perturbed, supporting Hypothesis 4.

\subsection{Discussion and Summary}

For the experiments with dihedral groups, we can see in Figures \ref{fig:symgroup_val_scores}(b) and (c) that the \(SQG\) metrics for the drawing of the dodecahedral graph displaying dihedral symmetry of order 2 are lower than that of the drawing of the cuboctahedral graph displaying dihedral symmetry of order 2. This is due to the difference in the maximum order of automorphism in the input, which is 10 for the dodecahedral graph and 6 for the cuboctahedral graph. Due to the \(SQ\) metrics for each automorphism being weighted depending on the sizes of their orbits when computing \(SQG\), this causes the dihedral symmetry of order 2 to have less relative weight in the dodecahedral graph case than the cuboctahedral graph case. Therefore, this experiment supports the usage of the weightings in our \(SQG\) formula (Equation \ref{eq:sqg}).

The results with perturbation experiments, such as with the example in Figure \ref{fig:c12x3_subrotdist}, show that not only that the computed scores decrease when the drawing is deformed such that fewer orbits are displayed symmetrically, but also that the metric computes lower scores when the deformed drawing is closer to a symmetric drawing that displays a lower order of symmetry. This can be seen in the difference in the scores computed for \(C12D_1\) and \(C6D_1\), where the drawings were perturbed from a symmetric drawing whose largest rotational symmetry is of order 12 and 6 respectively, in Figure \ref{fig:symgroup_val_scores}(d) to the scores obtained for \(D4\) and \(D5\), showing rotational symmetries of order 3 and 2 respectively, in Figure \ref{fig:symgroup_val_scores}(a).

\textit{In summary, our experiments have supported Hypotheses 3 and 4, validating the effectiveness of the \(SQG\) metric in capturing the difference in quality of drawings displaying different subsets of an automorphism group of a graph, as well as in capturing the difference in quality when such drawings are perturbed to produce asymmetric drawings.}

\section{Experiment 3: Layout Comparison Experiments}

\subsection{Experiment Design}

After validating the usage of the \(SQG\) metric to compute a quantitative score measuring how well a drawing displays a group of automorphisms of a graph, we conduct experiments comparing a number of different graph layout algorithms. We select the following automatic layouts to be compared: Fruchterman-Reingold (FR)~\cite{fruchterman1991graph}, Stress Majorization~\cite{gansner2004graph}, Pivot MDS~\cite{brandes2006eigensolver}, spectral, and Tutte. We also produce drawings using the \textit{Concentric Circles} layout~\cite{abelson2007geometric}, where we create concentric circles according to the number of orbits of a selected automorphism, assign each orbit to a circle, and place the vertices belonging to the orbit in a regular convex polygon position around the circle.

We perform the experiments using the following steps:

\begin{enumerate}
    \item \label{step:drawcomp} We select a symmetric graph \(G\) and draw it using all of the selected layout algorithms.
    \item We define a group of automorphisms \(A\) of \(G\) as the input.
    \item We compute the \(SQG\) metrics for the drawings produced by each layout algorithm with \(A\) as the input automorphism group.
\end{enumerate}

In Step \ref{step:drawcomp}, we generate one drawing each using each layout algorithm, except for FR, where we generate five drawings per graph, each time starting from a random initial layout. This is due to the non-deterministic nature of FR, compared to the other selected layout algorithms. We then average the \(SQG\) metrics computed for all the FR-generated drawings to obtain a value that can be compared to those computed for other layout algorithms.

Based on how the Concentric Circles layout places vertices in regular convex polygon positions along concentric circles, we expect that this layout will perform well in displaying automorphisms as geometric symmetry. We also expect that the Tutte layout, due to being designed to display planar graphs, will perform well in displaying \textit{planar automorphisms}, which are automorphisms that can be displayed in planar graph drawings. We formulate the following hypothesis for this experiment:

\begin{itemize}
    \item Hypothesis 5: The Concentric Circle layout will always attain \(SQG\) metrics of 1 and Tutte will always display planar automorphisms when the graph has planar automorphisms.
\end{itemize}

\subsection{Layout Comparison Results}

Figure \ref{fig:dodecahedral_comp} shows an example of the layout comparison experiment using the dodecahedral graph. In this experiment, we use the \(SQG\) metric with the input dihedral groups of order 10, 5, and 2. Figure \ref{fig:dodecahedral_comp_scores} displays the symmetry quality scores computed for this experiment. Concentric Circles obtains a metric score of 1, supporting Hypothesis 5. Stress Majorization also obtains a metric score of 1, and behind the two, Tutte obtains the third highest metric score.

A similar result is shown in the examples using the tesseract graph in Figures \ref{fig:tesseract_comp} and \ref{fig:tesseract_comp_scores}, where we take as input dihedral groups of order 8, 4, and 2. Concentric Circles and Stress Majorization again obtains \(SQG\) values of 1, with Tutte the third highest.

\begin{figure}[H]
    \centering
    \begin{tabular}{|c|c|c|c|c|c|}
    \hline
     Stress Maj. & Pivot MDS \\ \hline
     \includegraphics[width=0.4\columnwidth]{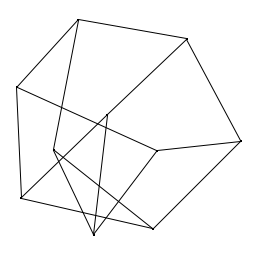} & \includegraphics[width=0.4\columnwidth]{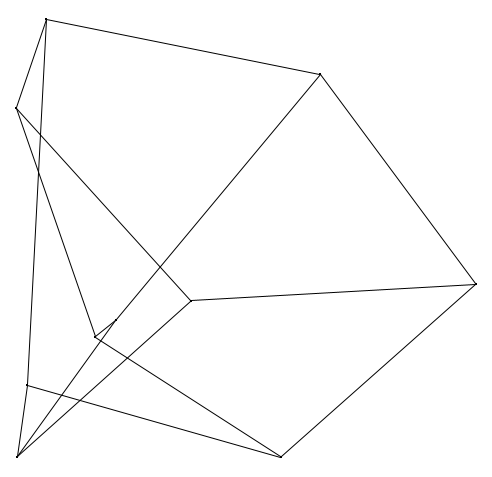} \\ \hline
     Spectral & FR \\ \hline
     \includegraphics[width=0.4\columnwidth]{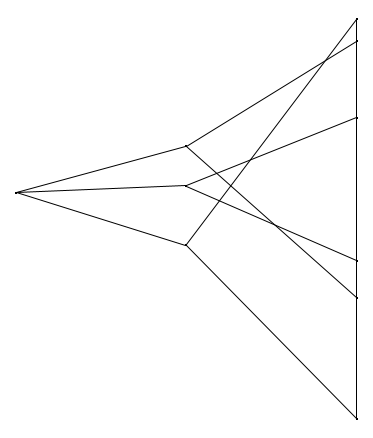} & \includegraphics[width=0.4\columnwidth]{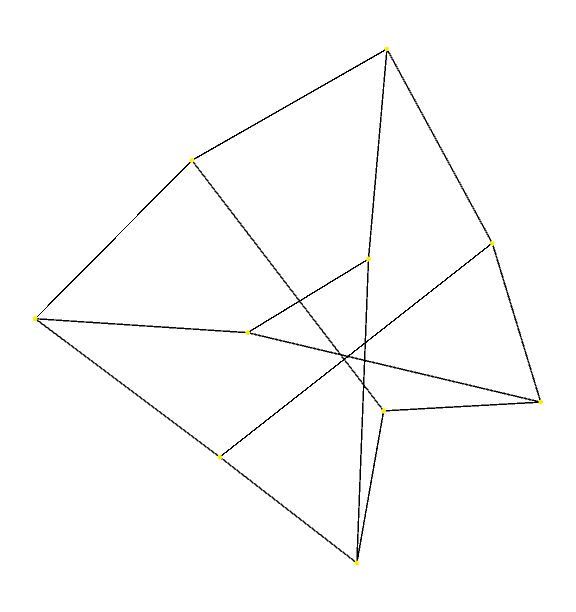}\\ \hline
     Tutte & Concentric Circles \\ \hline
     \includegraphics[width=0.4\columnwidth]{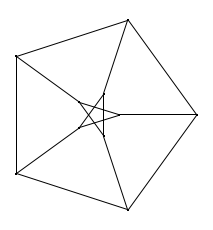} & \includegraphics[width=0.4\columnwidth]{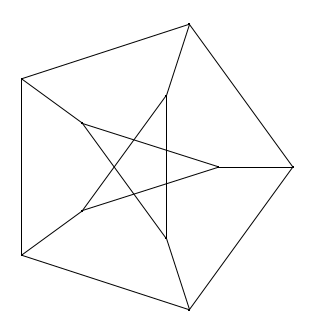} \\ \hline
    \end{tabular}
    \caption{Layout comparison experiment with the Petersen graph, with input dihedral group of order 5. Aside from concentric circles, Tutte is the only layout that perfectly displays the automorphism group as symmetries.}
    \label{fig:petersen_comp}
\end{figure}

\begin{figure}[H]
    \centering
    \includegraphics[width=0.95\columnwidth]{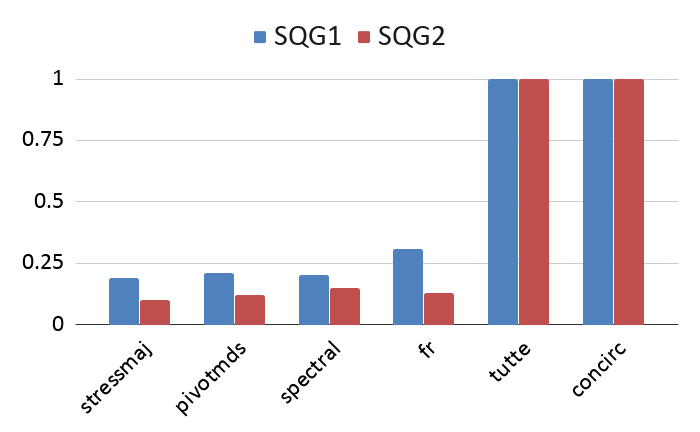}
    \caption{\(SQG\) metrics computed for the layout comparison experiment with the Petersen graph (Figure \ref{fig:petersen_comp}). Concentric Circles and Tutte obtain scores of 1, while all other layouts obtain scores of less than 0.3.}
    \label{fig:petersen_comp_scores}
\end{figure}

Figure \ref{fig:petersen_comp} shows another example of the layout comparison experiment, taking as input dihedral group of order 5 on the Petersen graph. Figure \ref{fig:petersen_comp_scores} shows the \(SQG_1\) and \(SQG_2\) computed for this experiment. Although the automorphism is not planar, Tutte obtains \(SQG\) of one, the same as Concentric Circles, while all other layouts obtain \(SQG\) of lower than 0.3.

\subsection{Discussion and Summary}

\begin{figure}[H]
    \centering
    \includegraphics[width=0.9\columnwidth]{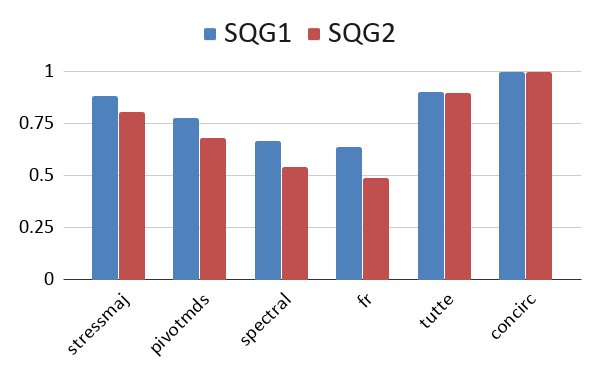}
    \caption{Average \(SQG\) metrics computed for all nine layout comparison graphs. Concentric Circles, Tutte, and Stress Majorization obtain the three highest scores.}
    \label{fig:layoutcomp_avg}
\end{figure}

The results for layout comparisons, as summarized in the averages over nine graphs shown in Figure \ref{fig:layoutcomp_avg}, show that the Concentric Circles, Tutte, and Stress Majorization layouts obtains the top 3 values for the \(SQG\) metric. The results for Concentric Circles supports Hypothesis 5.

\begin{figure}[H]
    \centering
    \begin{tabular}{|c|c|c|c|c|c|}
    \hline
     Stress Maj. & Pivot MDS \\ \hline
     \includegraphics[width=0.4\columnwidth]{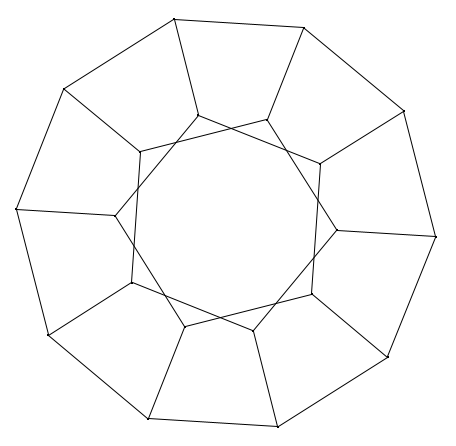} & \includegraphics[width=0.4\columnwidth]{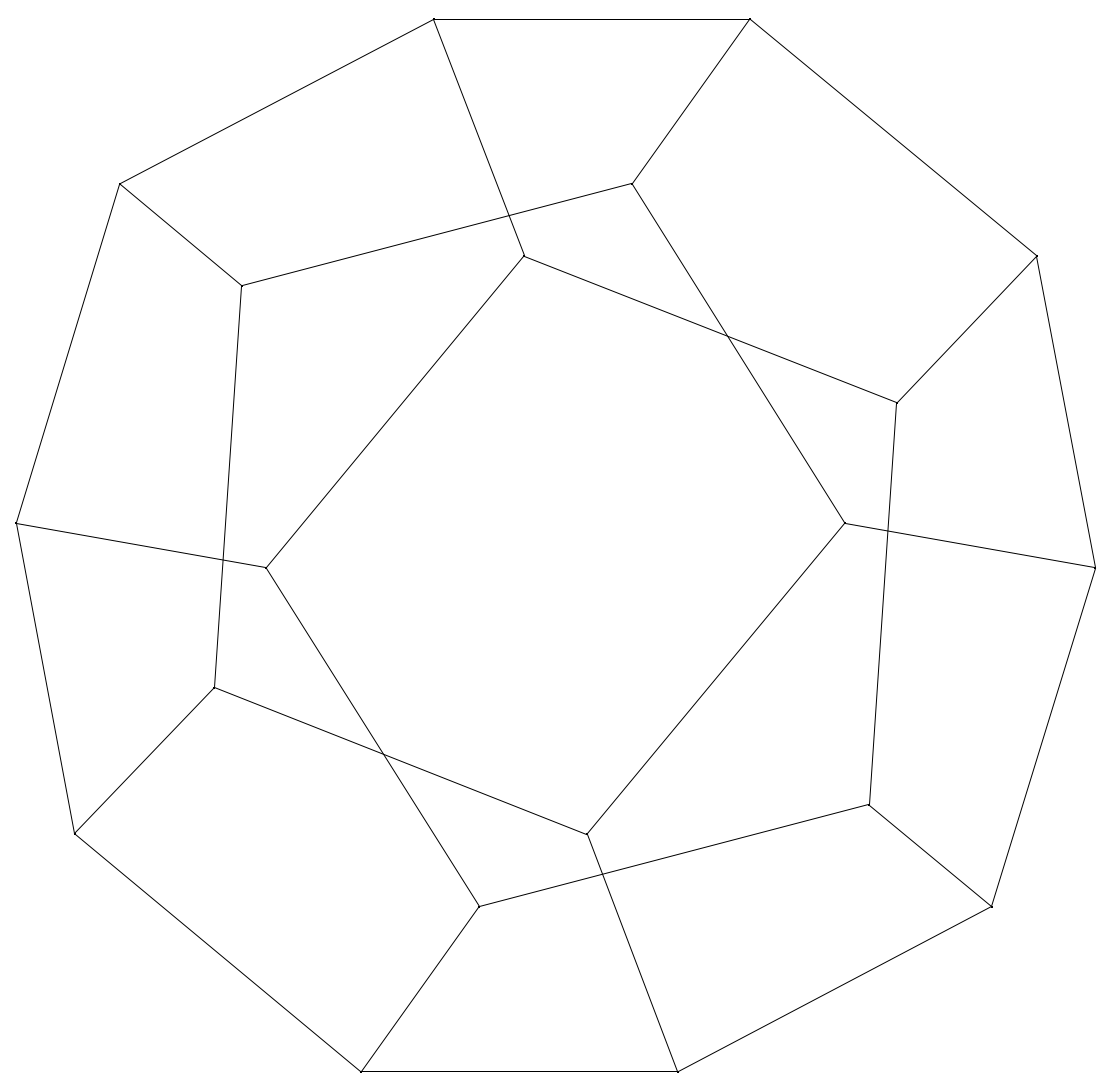} \\ \hline
     Spectral & FR \\ \hline
      \includegraphics[width=0.4\columnwidth]{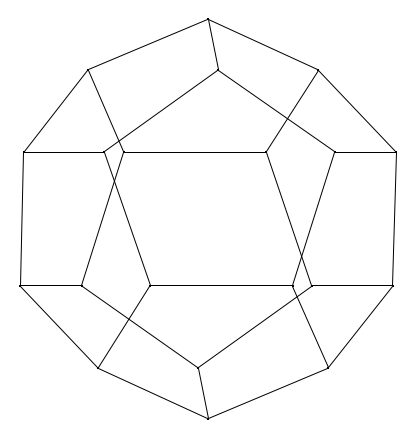} & \includegraphics[width=0.4\columnwidth]{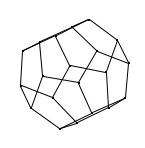}\\ \hline
     Tutte & Concentric Circles \\ \hline
      \includegraphics[width=0.4\columnwidth]{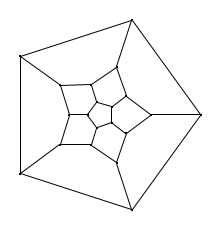} & \includegraphics[width=0.4\columnwidth]{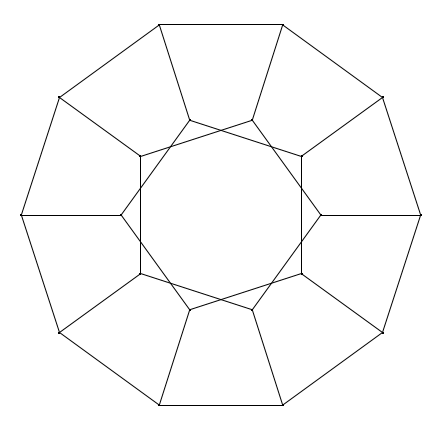} \\ \hline
    \end{tabular}
    \caption{Layout comparison experiment with the dodecahedral graph, detecting dihedral symmetries of order 10, 5, and 2. Among automatic layouts, Stress Majorization realizes all of the input automorphisms, while Tutte displays the largest planar automorphism as dihedral symmetry of order 5.}
    \label{fig:dodecahedral_comp}
\end{figure}

\begin{figure}[H]
    \centering
    \includegraphics[width=0.95\columnwidth]{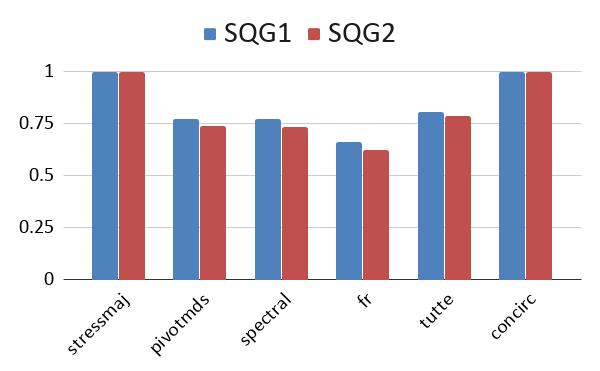}
    \caption{\(SQG\) metrics computed for the layout comparison experiment with the dodecahedral graph (Figure \ref{fig:dodecahedral_comp}). Concentric Circles and Stress Majorization obtain scores of 1, while Tutte, which displays a planar automorphism of a lower order, obtains lower scores than the two.}
    \label{fig:dodecahedral_comp_scores}
\end{figure}

\begin{figure}[H]
    \centering
    \begin{tabular}{|c|c|c|c|c|c|}
    \hline
     Stress Maj. & Pivot MDS \\ \hline
     \includegraphics[width=0.4\columnwidth]{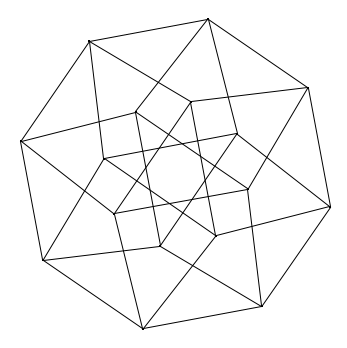} & \includegraphics[width=0.4\columnwidth]{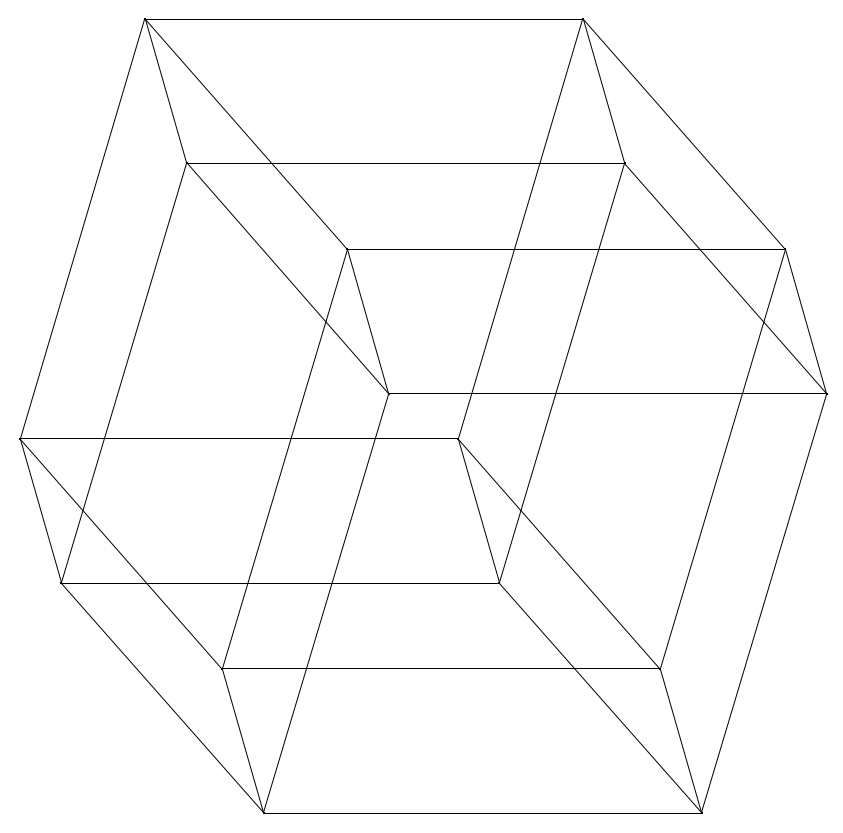} \\ \hline
     Spectral & FR \\ \hline
     \includegraphics[width=0.4\columnwidth]{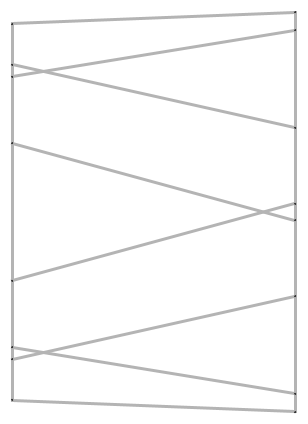} & \includegraphics[width=0.4\columnwidth]{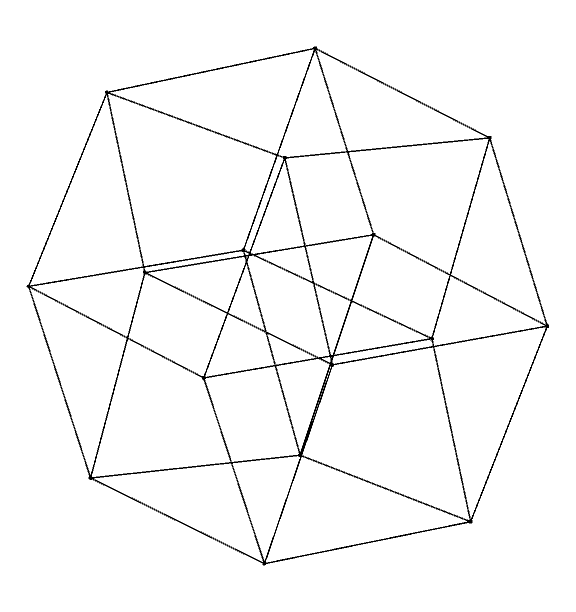}\\ \hline
     Tutte & Concentric Circles \\ \hline
      \includegraphics[width=0.4\columnwidth]{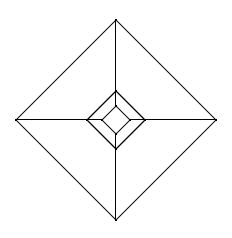} & \includegraphics[width=0.4\columnwidth]{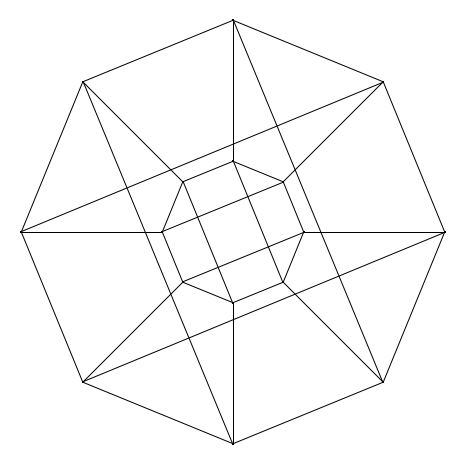} \\ \hline
    \end{tabular}
    \caption{Layout comparison experiment with the tesseract graph, detecting dihedral symmetries of order 8, 4, and 2. Similar to the dodecahedral graph in Figure \ref{fig:dodecahedral_comp}, Stress Majorization displays the highest order of automorphism (order 8) while Tutte realizes the highest order of planar automorphism with dihedral symmetry of order 4.}
    \label{fig:tesseract_comp}
\end{figure}

\begin{figure}[H]
    \centering
    \includegraphics[width=0.95\columnwidth]{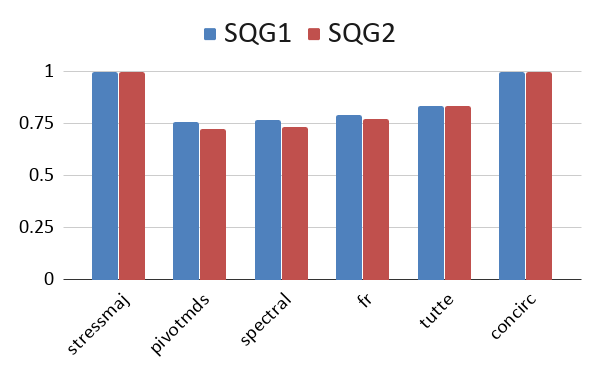}
    \caption{\(SQG\) metrics computed for the layout comparison experiment with the tesseract graph (Figure \ref{fig:tesseract_comp}). Concentric Circles and Stress Majorization, displaying the highest order of automorphism, obtains scores of 1.}
    \label{fig:tesseract_comp_scores}
\end{figure}

The results for the Tutte layout may arise from the nature of the layout algorithm, where given a set of vertices forming the outer face fixed as a regular convex polygon, all other vertices are placed at the barycenter of its neighbors. This results in a unique embedding that minimizes a linear system of equations, which could lead to its ability to capture automorphisms of the graph as symmetries. However, as the layout is designed for planar graphs, the results favor planar automorphisms. This causes it to score lower on examples such as the dodecahedral graph in Figure \ref{fig:dodecahedral_comp}, where it displays a dihedral automorphism of order 5, compared to Stress Majorization which displays the (non-planar) dihedral automorphism of order 10.

With Stress Majorization, it has been shown that it is possible to obtain symmetric layouts with low stress starting from the Concentric Circles layout~\cite{eades2000spring}. Unlike force-directed layouts such as FR, the set of minima for stress-based layouts is more limited, increasing the chances of producing a symmetric layout, which can explain its superior performance compared to FR.

\textit{In summary, our experiments have supported hypothesis 5 by showing that the concentric circles layout always obtains a score of 1 on \(SQG\) and that Tutte layout is always able to display the highest order of planar automorphism when a graph possesses planar automorphisms. We also observe that Stress Majorization performs better at displaying graph automorphisms as symmetries compared to the remaining tested layouts: FR, Pivot MDS, and Spectral.}

\section{Conclusion}

We have introduced a new quality metric to measure how faithfully a graph drawing visualizes the graph's automorphisms as symmetries, based on both Euclidean distance and mathematical group theory. The metrics, \(SQ\) and \(SQG\), are suited to detect symmetries corresponding to rotational, reflectional, and dihedral automorphisms.

We defined algorithms to compute the metrics, running in \(O(n \log n)\) time for the \(SQ\) version taking as input a single automorphism and in \(O(kn \log n)\) time for the \(SQG\) version taking as input an automorphism group, with \(n\) as the number of vertices and \(k\) as the size of the automorphism group.

Experiments have validated the effectiveness of both \(SQ\) and \(SQG\) in capturing the extent to which the drawing's symmetries reflect the automorphisms of the graph. We demonstrated the effectiveness of \(SQ\) in reflecting the extent of distortion of a drawing from exact symmetry, and the effectiveness of \(SQG\) in reflecting the difference in quality between drawings displaying different automorphism groups of a graph.

We have also compared layout algorithms using our metric, and in the process, confirm the effectiveness of the concentric circles layout in displaying a graph's automorphisms as symmetries and similarly for the Tutte layout for planar automorphisms.

Future work may include extending the metric to graph drawings in three dimensions.




\bibliographystyle{abbrv-doi}
\bibliography{template}

\begin{thebibliography}{10}

\bibitem{abelson2007geometric}
D.~Abelson, S.-H. Hong, and D.~E. Taylor.
\newblock Geometric automorphism groups of graphs.
\newblock {\em Discrete Applied Mathematics}, 155(17):2211--2226, 2007.

\bibitem{alt1988congruence}
H.~Alt, K.~Mehlhorn, H.~Wagener, and E.~Welzl.
\newblock Congruence, similarity, and symmetries of geometric objects.
\newblock {\em Discrete \& Computational Geometry}, 3(3):237--256, 1988.

\bibitem{battista1998graph}
G.~D. Battista, P.~Eades, R.~Tamassia, and I.~G. Tollis.
\newblock {\em Graph drawing: algorithms for the visualization of graphs}.
\newblock Prentice Hall PTR, 1998.

\bibitem{brandes2006eigensolver}
U.~Brandes and C.~Pich.
\newblock Eigensolver methods for progressive multidimensional scaling of large
  data.
\newblock In {\em International Symposium on Graph Drawing}, pp. 42--53.
  Springer, 2006.

\bibitem{buchheim2003}
C.~Buchheim and M.~J{\"u}nger.
\newblock Detecting symmetries by branch \& cut.
\newblock {\em Mathematical programming}, 98(1-3):369--384, 2003.

\bibitem{coxeter1983my}
H.~Coxeter.
\newblock My graph.
\newblock {\em Proceedings of the London Mathematical Society}, 3(1):117--136,
  1983.

\bibitem{defraysseix1999heuristic}
H.~De~Fraysseix.
\newblock An heuristic for graph symmetry detection.
\newblock In {\em International Symposium on Graph Drawing}, pp. 276--285.
  Springer, 1999.

\bibitem{eades2013detection}
P.~Eades and S.~Hong.
\newblock Detection and display of symmetries.
\newblock {\em Handbook of Graph Drawing and Visualisation. CRC Press, Boca
  Raton, FL, to appear}, 2013.

\bibitem{eades2000spring}
P.~Eades and X.~Lin.
\newblock Spring algorithms and symmetry.
\newblock {\em Theoretical Computer Science}, 240(2):379--405, 2000.

\bibitem{fruchterman1991graph}
T.~M.~J. Fruchterman and E.~M. Reingold.
\newblock Graph drawing by force-directed placement.
\newblock {\em Software: Practice and Experience}, 21(11):1129--1164, 1991.
  doi: {{%
10\hspace{.1pt}\discretionary{.}{%
}{.}\hspace{.4pt}1002\discretionary{/}{%
}{/}spe\hspace{.1pt}\discretionary{.}{%
}{.}\hspace{.4pt}4380211102}}


\bibitem{gansner2004graph}
E.~R. Gansner, Y.~Koren, and S.~North.
\newblock Graph drawing by stress majorization.
\newblock In J.~Pach, ed., {\em International Symposium on Graph Drawing}, pp.
  239--250. Springer Berlin Heidelberg, 2004.

\bibitem{hong2003symmetric}
S.-H. Hong and P.~Eades.
\newblock Symmetric layout of disconnected graphs.
\newblock In {\em International Symposium on Algorithms and Computation}, pp.
  405--414. Springer, 2003.

\bibitem{hong2005drawing}
S.-H. Hong and P.~Eades.
\newblock Drawing planar graphs symmetrically, ii: Biconnected planar graphs.
\newblock {\em Algorithmica}, 42(2):159--197, 2005.

\bibitem{hong2006drawing}
S.-H. Hong and P.~Eades.
\newblock Drawing planar graphs symmetrically, iii: Oneconnected planar graphs.
\newblock {\em Algorithmica}, 44(1):67--100, 2006.

\bibitem{hong2000drawing}
S.-H. Hong, P.~Eades, and S.-H. Lee.
\newblock Drawing series parallel digraphs symmetrically.
\newblock {\em Computational Geometry}, 17(3-4):165--188, 2000.

\bibitem{hong2006linear}
S.-H. Hong, B.~McKay, and P.~Eades.
\newblock A linear time algorithm for constructing maximally symmetric straight
  line drawings of triconnected planar graphs.
\newblock {\em Discrete \& Computational Geometry}, 36(2):283--311, 2006.

\bibitem{klapaukh2018}
R.~Klapaukh, S.~Marshall, and D.~Pearce.
\newblock A symmetry metric for graphs and line diagrams.
\newblock In P.~Chapman, G.~Stapleton, A.~Moktefi, S.~Perez-Kriz, and
  F.~Bellucci, eds., {\em Diagrammatic Representation and Inference}, pp.
  739--742. Springer International Publishing, Cham, 2018.

\bibitem{lipton1985method}
R.~J. Lipton, S.~C. North, and J.~S. Sandberg.
\newblock A method for drawing graphs.
\newblock In {\em Proceedings of the first annual symposium on Computational
  geometry}, pp. 153--160. ACM, 1985.

\bibitem{lubiw1981}
A.~Lubiw.
\newblock Some np-complete problems similar to graph isomorphism.
\newblock {\em SIAM Journal on Computing}, 10(1):11--21, 1981.

\bibitem{deluca2019}
F.~D. Luca, M.~I. Hossain, and S.~G. Kobourov.
\newblock Symmetry detection and classification in drawings of graphs.
\newblock {\em CoRR}, abs/1907.01004, 2019.

\bibitem{manning1988fast}
J.~Manning and M.~J. Atallah.
\newblock Fast detection and display of symmetry in trees.
\newblock {\em Congressus Numerantium}, 64:159--169, 1988.

\bibitem{manning1992fast}
J.~Manning and M.~J. Atallah.
\newblock Fast detection and display of symmetry in outerplanar graphs.
\newblock {\em Discrete Applied Mathematics}, 39(1):13--35, 1992.

\bibitem{manning1992geometric}
J.~B. Manning.
\newblock {\em Geometric Symmetry in Graphs}.
\newblock PhD thesis, West Lafayette, IN, USA, 1991.
\newblock UMI Order No. GAX91-16430.

\bibitem{purchase2002metrics}
H.~C. Purchase.
\newblock Metrics for graph drawing aesthetics.
\newblock {\em Journal of Visual Languages \& Computing}, 13(5):501--516, 2002.

\bibitem{wolter1985optimal}
J.~D. Wolter, T.~C. Woo, and R.~A. Volz.
\newblock Optimal algorithms for symmetry detection in two and three
  dimensions.
\newblock {\em The Visual Computer}, 1(1):37--48, 1985.

\bibitem{zabrodsky1995continuous}
H.~Zabrodsky and D.~Avnir.
\newblock Continuous symmetry measures. 4. chirality.
\newblock {\em Journal of the American Chemical Society}, 117(1):462--473,
  1995.

\bibitem{zabrodsky1995symmetry}
H.~Zabrodsky, S.~Peleg, and D.~Avnir.
\newblock Symmetry as a continuous feature.
\newblock {\em IEEE Transactions on Pattern Analysis \& Machine Intelligence},
  (12):1154--1166, 1995.

\end{thebibliography}
\end{document}